\pdfoutput=1
\pdfoutput=1
\pdfoutput=1
\pdfoutput=1
\pdfoutput=1

\documentclass{article}

\usepackage{arxiv}

\usepackage[utf8]{inputenc} 
\usepackage[T1]{fontenc}    
\usepackage{hyperref}       
\usepackage{url}            
\usepackage{booktabs}       
\usepackage{amsfonts}       
\usepackage{nicefrac}       
\usepackage{microtype}      

\usepackage{natbib}
\usepackage{framed,multirow}

\usepackage{xcolor}
\usepackage{csquotes}
\usepackage{subfigure}
\usepackage{wrapfig}
\usepackage{amsmath}
\usepackage{amsbsy}
\usepackage{psfrag}
\usepackage{pstool}
\usepackage{graphicx}
\usepackage{algorithm}
\usepackage{algpseudocode}
\usepackage{lineno,hyperref}
\modulolinenumbers[5]
\usepackage{tikz}
\usetikzlibrary{plotmarks}

\usepackage{ulem}

\makeatletter

\def\BState{\State\hskip-\ALG@thistlm}

\makeatother

\newlength\myindent
\title{A fast GPU Monte Carlo Radiative Heat Transfer Implementation for Coupling with Direct Numerical Simulation}

\author{
  \textbf{Simone Silvestri} \thanks{email: \texttt{s.silvestri@tudelft.nl}.} \hspace{0.2cm} \normalfont{and}  \hspace{0.2cm} \textbf{Rene Pecnik} \thanks{email: \texttt{r.pecnik@tudelft.nl}.} \\
  Department of Process and Energy\\
  Delft University of Technology\\
  Delft, The Netherlands, 2628CB \\
}

\begin{document}
\maketitle

\begin{abstract}
We implemented a fast Reciprocal Monte Carlo algorithm, to accurately solve radiative heat transfer in turbulent flows of non-grey participating media that can be coupled to fully resolved turbulent flows, namely to Direct Numerical Simulation (DNS). The spectrally varying absorption coefficient is treated in a narrow-band fashion with a correlated-$k$ distribution. The implementation is verified with analytical solutions and validated with results from literature and line-by-line Monte Carlo computations. 
The method is implemented on GPU with a thorough attention to memory transfer and computational efficiency. The bottlenecks that dominate the  computational expenses are addressed and several techniques are proposed to optimize the GPU execution. By implementing the proposed algorithmic accelerations, a speed-up of up to 3 orders of magnitude can be achieved, while maintaining the same accuracy. 
\end{abstract}


\definecolor{red1}{HTML}{EEEEFF} 
\definecolor{beige1}{HTML}{F4F3DF}

\section{Introduction}
Modeling radiative heat transfer is a challenging task due to the numerical complexity and the associated computational costs \cite{modestbook}. For example, radiative heat transfer is a six dimensional problem, which depends on spatial location, propagation direction and frequency of the electromagnetic wave. 
In addition, the calculation of radiative heat transfer poses a daunting challenge when it is coupled with convective and conductive heat transfer modes in 
turbulent flows. The computational cost of solving the radiative transfer equation makes it difficult to obtain an accurate description on how radiative transfer couples to a participating turbulent fluid flow. 
As a consequence, a complete view of the interplay between turbulence and radiation is missing. 

Recently, several studies have addressed the problem with the aid of simplifying assumptions to ease the computational burden. In particular, Sakurai et al. \cite{sakurai} used the Optically Thin Approximation (OTA) to study the influence of radiative effects in a horizontally buoyant turbulent channel flow. They noticed that large scale buoyant structures are destroyed by the presence of non-local radiative heat transfer. The OTA assumes the intensity to be independent of spatial position, leading to a constant incident radiation throughout the domain. This assumption greatly simplifies the description of radiative heat transfer. However, it does not allow the evaluation of incident radiative fluctuations and is, therefore, restricted to low values of absorption coefficient, $\kappa$, as demonstrated in Ref.~\cite{roekaerts}.

A common approximation employed in solving the radiative transfer equation (RTE) in DNS coupled simulations consists in neglecting the spectral dependency of $\kappa$ by assuming a grey gas. This assumption enables the efficient use of finite difference schemes, such as the discrete ordinates method (DOM), which require a low computational effort and provide a high level of accuracy. Examples are given in Refs.~\citep{deshmukh, ghosh, ghosh2,  gupta, simone}, who 
have studied the influence of radiative heat transfer in turbulent flows using the grey gas approximation, coupled to either DNS or large eddy simulations (LES). These studies allow to investigate the effect of radiation on the turbulent temperature field and vice versa, to highlight and quantify the dissipative effect of radiative field fluctuations and the impact of Turbulence Radiation Interactions (TRI). However, these cases are highly idealized since radiation in real fluids is intrinsically non-grey.  

If a spectral description of radiative heat transfer is to be included, the state of art involves the use of a Monte Carlo (MC) method. Compared to the above mentioned RTE solution methods, the Monte Carlo method can be considered the most accurate and flexible. Its solution time increases mildly with problem complexity, allowing a detailed spectral description or the simulations of complex geometries, which are challenging with other methods such as DOM. 
To the authors knowledge, the first instance of a Monte Carlo method coupled with DNS is reported by Wu et al.~\cite{wu}. They developed a high resolution MC method, subsequently used by Deshmukh et al. \cite{deshmukh} to study TRI in a statistically one dimensional premixed combustion system. In their study, they noticed that absorption TRI intensifies with an increase of optical thickness, while emission TRI is always relevant in reactive flows. The calculations were performed on a $64^3$ mesh in a grey gas, but the use of a MC method coupled to DNS on finer grids and spectral medium was not investigated. On the other hand, more recently, Vicquelin et al. \cite{vicquelin} performed DNS of radiative channel flow using a MC code coupled to a narrow-band correlated-$k$ spectral description. They investigated the effect of different radiative budgets to verify the modification of first and second order temperature and velocity statistics. Nonetheless, the computational expenses of the MC method prohibited the solution of the radiative heat transfer on a full DNS grid, which required an intermediate interpolation step between flow and radiation solution. Overall, the computational expenses of the Monte Carlo solver limit the possibilities of accurately solving coupled radiative heat transfer and turbulence accurately. However, since the Monte Carlo method is \enquote{embarassingly parallelizable} (i.e., can be divided into a number of completely independent computations), it greatly benefits from the use of parallel architectures and in particular from the use of general purpose graphical processing units (GPGPU). 

The use of GPUs for computational sciences has become increasingly investigated, especially for large parallelizable problems that are more efficiently mapped on many GPU parallel multiprocessors \cite{owens}. The development of NVIDIA CUDA, a versatile GPU programming language, has further popularized GPUs as accelerators alongside CPU computational clusters \cite{nvidia}. Several examples of GPU codes are available up to date, ranging from machine learning \cite{machine} to imaging \cite{imaging} and computational biology \cite{nobile}. Likewise, in the fluid mechanics field, Khajeh et al. \cite{khajeh} and Salvadore et al. \cite{salvadore} have been porting a Navier-Stokes solver on GPUs obtaining speedups up to 22 $\times$. Additionally, many Monte Carlo codes have been developed on graphical processing units for many diverse fields and applications such as finance \cite{finance} and molecular dynamics \cite{liang}. On the other hand, to the authors knowledge, the only MC method applied to the solution of thermal radiation implemented on GPU was developed by Humphrey et al.~\cite{humphrey} for grey gas applications. Their code showed excellent scaling capabilities up to 16834 GPUs, proving the feasibility of the GPU MC concept for thermal radiation. 

Nonetheless, porting an application to GPU requires the exposure of the parallel portion of the application and algorithmic optimizations to improve the efficiency on a GPU architecture. Therefore, the objective of this work is to develop an optimized GPU Monte Carlo implementation, which can enable a fast and accurate solution of radiative heat transfer in largely fluctuating temperature fields typical of turbulent flows. We will include the spectral description of the absorption coefficient to have a complete and flexible solver. All the challenges involved in implementing an efficient GPU application are addressed in order to reduce the computational time and to improve the scaling with problem size.  

\section{The Monte Carlo method}
\label{numerics}
In this section the details of the Monte Carlo methods are outlined for the sake of completeness. Within a domain containing a non grey absorbing and emitting medium, the radiative power emitted by cell $i$ and absorbed within cell $j$ is expressed, as in Tesse et al. \cite{tesse}, by
\begin{linenomath*}\begin{equation}
Q_{i \rightarrow j}^R = \int_0^\infty \kappa_{\nu}(T_i) {I_b}_{\nu}(T_i)  \int_{V_i} \int_{4\pi} \sum_{m=1}^{N_c}  \tau_{\nu} (i \rightarrow j,m) \left[ \int_{0}^{l_{j,m}} \kappa_{\nu}(T_j) e^{ - \kappa_{\nu}(T_j) s_{j,m}} ds_{j,m}    \right] d \Omega d V_i d \nu ,
\label{MCea}
\end{equation}\end{linenomath*}
where $\nu$ is the wavenumber, $\kappa_{\nu}$ is the spectral absorption coefficient, $\tau_{\nu}$ is the spectral transmissivity from cell $i$ to the boundary of cell $j$ following the path $m$, $N_c$ is the number of paths that, from cell $i$, cross cell $j$, and $l_{j,m}$ is the distance travelled in cell $j$ along the propagation direction. The volume integral $V_j$, as given in ref.~\cite{cherkaoui} has been replaced by the integration over the solid angle $\Omega$ and the path length $s_{j,m}$ as done in ref.~\cite{tesse}. The integral in the square brackets represents the absorption within cell $j$, following path $m$. The analytical solution, considering cell $j$ is isothermal and homogeneous, is 
\begin{linenomath*}\begin{equation}
{\alpha_{\nu}}_{j,m} = 1 - e^{ - \kappa_{\nu}(T_j) l_{j,m}}.
\label{alpha} 
\end{equation}\end{linenomath*}
The spectral transmissivity $\tau_{\nu}(i \rightarrow j,m)$ is the result of the absorption by the finite volumes and surfaces crossed by path $m$, and can be calculated as 
\begin{linenomath*}\begin{equation}
\tau_{\nu}(i \rightarrow j,m ) = \prod_{k=i}^{j-1} (1-{\alpha_{\nu}}_{k,m} )\times \prod_{c=1}^{N_r} (1 - \varepsilon_{w}) \ , 
\label{tau}
\end{equation}\end{linenomath*}
where $\varepsilon_{w}$ is the wall emissivity and $N_r$ is the number of wall reflections that occurred for path $m$.

The Monte Carlo method consists in a statistical estimation of the integrals in equation (\ref{MCea}) using a large number of samples that represent different paths and wavelengths. In particular, it is possible to develop probability distribution functions defined as
\begin{linenomath*}\begin{equation}
f_V = \frac{1}{V_i} \ , \ \ \ \ f_{\theta} = \frac{\sin{\theta}}{ 2 }  \ , \ \ \ \ f_{\phi} = \frac{1} {2 \pi} \ , \ \ \ \ f_{\nu} = \frac{ \pi \kappa_{\nu}(T_{i}) {I_b}_{\nu}(T_{i})}{{\kappa_p}(T_i) \sigma T_i^4} \ ,
\label{prob}
\end{equation}\end{linenomath*}
where ${\kappa_p}(T_i)$ is the Planck mean absorption coefficient of cell $i$, while $\theta$ and $\phi$ are the polar and azimuthal angles, respectively, with $d \Omega = \sin \theta d\theta d \phi$. Substituting the probability distribution functions in equation (\ref{MCea}) leads to 
\begin{linenomath*}\begin{equation}
Q_{i \rightarrow j}^R =  {Q}^{R,e}(T_i) \int_0^\infty f_{\nu}  \int_{V_i} f_V \int_{0}^{2\pi} \ f_{\phi} \int_0^{\pi} f_{\theta} \  A_{\nu,m,i \rightarrow j} \ d \theta \ d \phi \ d V_i \ d \nu  \ ,
\label{MCint}
\end{equation}\end{linenomath*}
where ${Q}^{R,e}(T_i)$ and $A_{\nu,m,i \rightarrow j}$ are the total radiative power emitted by cell $i$ and the spectral energy fraction emitted by cell $i$ and absorbed in cell $j$ through path $m$, respectively. These are calculated using 
\begin{linenomath*}\begin{equation}
Q^{R,e}(T_i)=  4  V_i {\kappa_p}(T_i) \sigma T_i^4 , 
\end{equation}\end{linenomath*}
\begin{linenomath*}\begin{equation}
A_{\nu,m,i \rightarrow j} =\sum_{m=1}^{N_c}  \tau_{\nu} (i \rightarrow j,m){\alpha_{\nu}}_{j,m} . 
\end{equation}\end{linenomath*}
A statistical estimation of the integrals in equation (\ref{MCint}) involves launching several samples, referred hereafter as \enquote{rays} with properties that are sampled from probability density functions as given in equation~(\ref{prob}).\\
The resulting discretized equation has then the form 
\begin{linenomath*}\begin{equation}
\widetilde{Q_{i \rightarrow j}^R} = \frac{{Q}^{R,e}(T_i)}{N_r} \sum_{r = 1}^{N_r} A_{r, i \rightarrow j} \ .
\label{Pow}
\end{equation}\end{linenomath*}
The tilde $\sim$ denotes a statistical estimator and the subscript $r$ indicates a ray, characterized by its wavenumber $\nu$, and direction angles $\theta$ and $\phi$ (defining the path variable $m$), which are calculated inverting the following relations 
\begin{linenomath*}\begin{equation}
\begin{split}
R_{\nu} & = \int_0^{\nu} f_{\nu^\prime}(T) d \nu^\prime = \frac{\pi \int_0^{\nu} \kappa_{\nu^\prime}(T) {I_b}_{\nu^\prime}(T) d \nu^\prime}{\kappa_p(T) \sigma T^4} \ , \\
 R_{\theta} &  = \int_0^{\theta} f_{\theta^\prime} d \theta^\prime = \frac{1 - \cos{\theta}} {2}  \ , \\
 R_{\phi} & = \int_0^{\phi}f_{\phi^\prime} d \phi^\prime =  \frac{\phi} {2 \pi} \ .
\end{split}
\label{inversion}
\end{equation}\end{linenomath*}
$R_{\nu}$, $R_{\theta}$ and $R_{\phi}$ are random numbers sampled from a uniform probability distribution function between $0$ and $1$.\\
In a reciprocal Monte Carlo formulation, both emitted and absorbed power are statistically estimated as
\begin{linenomath*}\begin{equation}
Q^R_{i,RM} = \underbrace{\sum_{j = 1}^{N_v + N_s} \widetilde{Q_{i \rightarrow j}^R}}_{Q^{R,e}_i} - \underbrace{\sum_{j = 1}^{N_v + N_s} \widetilde{Q_{j \rightarrow i}^R}}_{Q^{R,a}_i} \ ,
\label{FM}
\end{equation}\end{linenomath*}
where $N_v + N_s$ are the numbers of volume and surfaces that interact with cell $i$. The reciprocal formulation employes the following principle 
\begin{linenomath*}\begin{equation}
\frac{Q^R_{i \rightarrow j, \nu}}{{I_b}_{\nu}(T_i)} = \frac{Q^R_{j \rightarrow i, \nu}}{{I_b}_{\nu}(T_j)} \ ,
\end{equation}\end{linenomath*}
to automatically satisfy the reciprocity condition. As a consequence, the above formulation avoids problems of large variance in case of low temperature gradients (i.e. non reactive flows) or high optical thickness that are typical of a forward Monte Carlo method.   
Depending on the estimated quantity, it is possible to distinguish between two reciprocity Monte Carlo formulations~\cite{tesse}. These are, the Absorption-based Reciprocity Monte Carlo (ARMC) which estimates the absorbed power, and the Emission-based Reciprocity Monte Carlo (ERMC) which estimates the emitted power. While ARMC results in a lower variance in low temperatures zones, characterized by relevant absorption, ERMC is more accurate in the high temperature regions that are dominated by emission. The advantage of ERMC is that $Q^R$ in $i$ is calculated by the emission of the cell, requiring only the computation of the rays leaving the cell itself. The corresponding relation of an ERMC formulation is given as 
\begin{linenomath*}\begin{equation} 
Q^R_{i,ERMC} = \sum_{j = 1}^{N_v + N_s} \widetilde{Q_{i \rightarrow j}^R} \cdot \left( 1 - \frac{{I_b}_{\nu}(T_j)}{{I_b}_{\nu}(T_i)} \right) \ .
\end{equation}\end{linenomath*}
Recently, Zhang et al. \cite{zhang} developed an optimized ERMC to reduce the variance in the low temperature regions. In the cold regions, $Q^R$ is dominated by the absorption of radiation which originates from hot zones. Nevertheless, an ERMC entails the estimation of absorption based on the emission of the cell itself. Consequently, the wavelength of emission in colder regions will be higher than the actual wavelength of the absorbed radiation that follows Wien's displacement law. This leads to a large variance in cold spots, which is characteristic for an ERMC based method. Therefore, ref.~\cite{zhang} proposed to sample the wavenumber from the maximum temperature, which corresponds to a larger emission in the domain using
\begin{linenomath*}\begin{equation}
f_{\nu} = \frac{ \pi \kappa_{\nu}(T_{max}) {I_b}_{\nu}(T_{max})}{{\kappa_p}(T_{max}) \sigma T_{max}^4} \ . 
\end{equation}\end{linenomath*}
As a result, equation (\ref{Pow}) has to be corrected with a prefactor $R_I$, resulting in 
\begin{linenomath*}\begin{equation}
\widetilde{Q_{i \rightarrow j}^R} = \frac{{Q}^{R,e}(T_{max})}{N_r} \sum_{r = 1}^{N_r} \bigg( \underbrace{\frac{\kappa_{\nu}(T_i){I_b}_{\nu}(T_i)}{\kappa_{\nu}(T_{max}){I_b}_{\nu}(T_{max})}}_{R_I} \bigg)~A_{r, i \rightarrow j} \ .
\label{finalMC}
\end{equation}\end{linenomath*}

\subsection{Spectral discretization}
In general, gas absorption spectra are characterized by discrete absorption lines, leading to a strong dependency on wavelength. In order to store the absorption coefficients and the probabilities associated with a line-by-line spectrum, comprised of more than a million spectral points, an excessive amount of memory is required. In addition, the high variability of the spectra translates in a lower convergence rate of the Monte-Carlo method. For this reason, we chose a narrow-band correlated-$k$ model to couple with the Monte Carlo solver \cite{soufiani}. 
The narrow-band method constitutes of an accurate spectral representation, comparable to a line-by-line description if enough pseudo-spectral points are considered, with significantly lower memory requirements. 
In addition it is naturally adaptable to a simple implementation of species transport and wavenumber-dependent scattering, in case multiphase flows are considered. The line-by-line spectrum of common gasses, for a wide range of temperatures and pressures, can be found in accurate online spectroscopy databases. For this study, the data from HITRAN 2012 \cite{hitran} is used to develop the narrow-band pseudo-spectral coefficients.
 
Since the narrow-band correlated-$k$ model divides the spectrum into narrow bands with assigned quadrature points, the wavenumber probability function in equation~(\ref{prob}) is discretized using two \enquote{discrete} probability functions, one for the narrow-band and the other one for the quadrature point. The two variables associated with the wavenumber of the photon bundle are thus a narrow band index $n$ and a quadrature point index $g$, 
\begin{linenomath*}\begin{equation}
\int_0^{\nu} f_{\nu^\prime} d \nu^\prime  \approx \sum_{n^\prime=1}^{n-1} f_{n^\prime} + f_{n} \cdot \sum_{g^\prime=1}^{g-1} f_{g^\prime}(n) \ ,
\end{equation}\end{linenomath*}
where
\begin{linenomath*}\begin{equation}
f_{n} = \frac{\pi \Delta \nu_n {I_b}_{n} \sum_{g^\prime=1}^{Nq} \omega_{g^\prime} k_{n,{g^\prime}}}{ \kappa_p \sigma T^4} \ , \ \ f_{g}(n) = \frac{\omega_g k_{n,g}  }{\sum_{g^\prime=1}^{Nq} \omega_{g^\prime} k_{n,g^\prime} } \ ,
\end{equation}\end{linenomath*}
and $\omega_g$ and $N_q$ are the Gaussian weights associated with point $g$ and the total number of quadrature points in a narrow band, respectively. Since the quadrature points in the narrow band all represent ideally the same wavenumber, the drawing of two independent random numbers is necessary in order to sample $n$ and $g$, 
\begin{linenomath*}\begin{equation}
R_n = \sum_{n^\prime = 1}^{n-1} f_{n^\prime} \ , \ \ R_g = \sum_{g^\prime = 1}^{g-1} f_{g^\prime}(n). 
\end{equation}\end{linenomath*}

\begin{figure}
\centering
\includegraphics[width=0.8\textwidth]{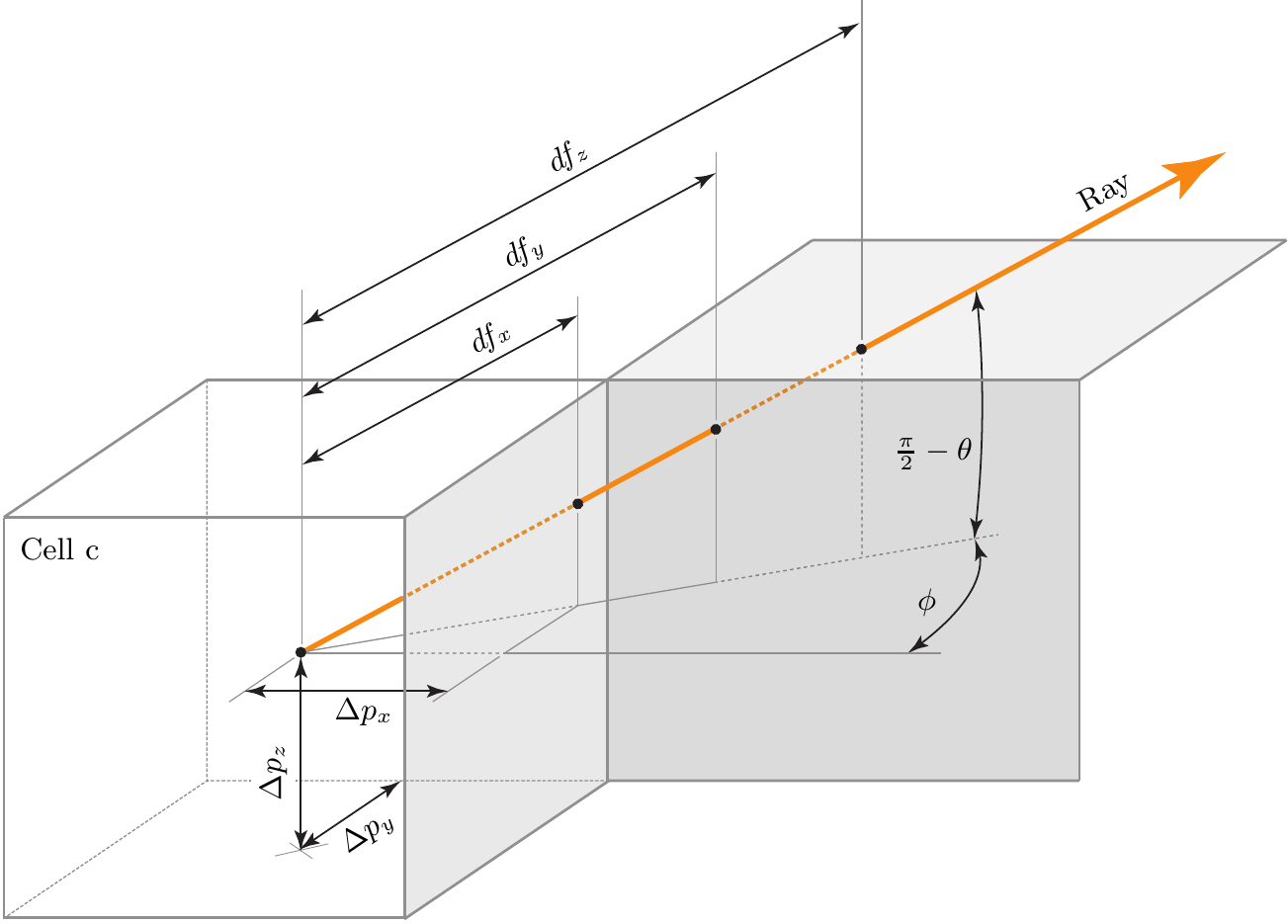}
\caption{Schematic displaying the marching ray procedure.}
\label{finite_volume}
\end{figure}
\subsection{Algorithm}
To ease the understanding of the GPU ERMC implementation, we first describe a standard CPU implementation in algorithm \ref{euclid}. The first loop (line 1) is performed over all finite volumes in the computational domain. Each finite volume is described by its index (i,j,k), and coordinates of the center and the surrounding faces. 
For each finite volume, a predefined number of rays (\texttt{numberOfRays}) are launched. The variable \texttt{ray} is a data structure that contains the current position (\texttt{pos}) of the ray and the index of the corresponding cell (\texttt{ind}), as well as the direction vector (\texttt{dir}) and the current transmissivity (\texttt{transmissivity}). The MC method mainly consists of two routines, the initialization (line 4) and the marching of the ray (line 16). In the first routine, the necessary random numbers are drawn and the properties of the ray are initialized accordingly. To accommodate a narrow-band correlated-$k$ description, two independent random numbers are drawn $R_{n}$ and $R_g$, which lead to two different indices \texttt{n} and \texttt{g} that specify the narrow band and the quadrature point within it. Marching the ray consists in finding the distances $\Delta p_x, \Delta p_y, \Delta p_z$, between the current location of the ray and the cell faces in direction \texttt{ray.dir}, specified by angles $\phi$ and $\theta$. The minimum distance, \texttt{ds}, determines which plane is crossed by the ray first. A schematic is displayed in figure~\ref{finite_volume}, for which the ray intersects the $x$-normal plane first, such that the minimum distance \texttt{ds} will be equal to $df_x$. The radiative power of the initial cell (\texttt{QR}) is then calculated in a reciprocal fashion. Furthermore, the new ray position and cell index are updated accordingly. If the transmissivity drops under a certain tolerance \texttt{tol} (line 17), the ray is terminated and the remaining energy is dumped into the initial cell (line 29). 
The on-the-fly calculation of the blackbody intensity from Planck's law is prohibitive due to the excessive computations involved. To overcome this issue, the blackbody intensity is precomputed for the narrow band wavelengths and discrete points in the required temperature range and then stored in an suitable 2D table. The functions \texttt{interpBlackbody} and \texttt{interpAbsorptionCoeff} (lines 12-14, 20 and 22) perform linear interpolations of the spectral blackbody intensity and the absorption coefficient from the corresponding tables, respectively. 
%
%
\begin{algorithm}[t]
\caption{ERMC CPU implementation}\label{euclid}
\begin{algorithmic}[1]
\scriptsize
\For{cell \textbf{in} Cells} \Comment{Loop over all finite volumes}
\State $\texttt{QE} \leftarrow 4\kappa_P(T_{max})\sigma T_{max}^4/$\texttt{numberOfRays} 
		\Comment{Cell emission $Q^{R,e}$ in equation (\ref{Pow}) }
\For{{ray} \textbf{in} Rays} 
		\Comment{Loop over rays}
\Procedure{Initialize}{}
\State $R_{\theta},R_{\phi},R_{n},R_{g} \leftarrow$ {\texttt{Rand}}(uniform distribution) 
		\Comment{Draw random numbers for angles and indices \texttt{n} and \texttt{g}} 
\State \texttt{ray.ind} $\leftarrow$ \texttt{cell.ind}
		\Comment{Initialize the ray with the cell index i,j, and k}
\State \texttt{ray.pos} $\leftarrow$ \texttt{cell.center} 
		\Comment{Initialize ray starting coordinates with cell center coordintes x, y, and z}
\State \texttt{ray.dir} $\leftarrow$ direction($R_{\theta}$, $R_{\phi}$) 
		\Comment{Find ray direction based on equation~(\ref{inversion})}
\State \texttt{ray.transmissivity} $\leftarrow$ 1.0
\State \texttt{indDir} $\leftarrow$ sign(\texttt{ray.dir})
		\Comment{Ray direction in terms of index i,j, and k}
\State $n,g$ $\leftarrow$ \texttt{findWavelength}{($R_{n}$, $R_g$)}
		\Comment{Binary search on CDF with $R_{n}$ and CDF(\texttt{n}) with $R_g$}
\State $Ib1 \leftarrow$ \texttt{interpBlackbody}($n$, temperature(\texttt{ray.ind})) 
		\Comment{blackbody intensity of initial cell c}
\State $R_I \leftarrow Ib1 \times$ \texttt{interpAbsorpCoeff}($n$, $g$, temperature(\texttt{ray.ind})) 
		\Comment{$R_I$ in equation~(\ref{finalMC})}
\State $R_I \leftarrow R_I$ / \texttt{interpBlackbody}(n, $T_{max}$) / \texttt{interpAbsorpCoeff}(n, g, $T_{max}$)
\EndProcedure
\Procedure{March}{}
\While{ $\texttt{ray.transmissivity} > tol$} 
\State $df \leftarrow \Delta p/\texttt{ray.dir}$ \Comment{Determine which face is crossed first (see figure~\ref{finite_volume})}
\State $ds \leftarrow min(df_x, df_y, df_z)$ 
		\Comment{Shortest distance is where ray crosses face}
\State $\kappa \leftarrow \texttt{interpAbsorpCoeff}(n, g, \textrm{temperature(\texttt{ray.ind})})$
\State $\alpha \leftarrow 1 - \text{exp}({-\kappa\times ds})$ 
		\Comment{equation~(\ref{alpha})}
\State $Ib2 \leftarrow \texttt{interpBlackbody}(n, \textrm{temperature(\texttt{ray.ind})})$
\State $\texttt{Absorption} \leftarrow \texttt{QE}\times \texttt{ray.transmissivity}\times \alpha \times ({Ib2}/{Ib1}-1) \times {R_I}$ 
		\Comment{equation~(\ref{finalMC})}
\State $\texttt{QR(cell.ind)} \leftarrow \texttt{QR(cell.ind)} - \texttt{Absorption}$  
		\Comment{radiative heat source of initial cell \texttt{c}}
\State $\texttt{ray.pos} \leftarrow \texttt{ray.pos} + ds\times \texttt{ray.dir}$
		\Comment{Update ray position}
\State $\texttt{ray.ind} \leftarrow \texttt{ray.ind} + \texttt{indDir}\times({ds}==[{df}_x,{df}_y,{df}_z])$
		\Comment{Update cell index depending on which face has been intersected}
\State $\texttt{ray.transmissivity} \leftarrow \texttt{ray.transmissivity}\times (1-\alpha)$ 
		\Comment{equation~(\ref{tau})}
\EndWhile
\State $\texttt{Absorption} \leftarrow \texttt{QE}\times \texttt{ray.transmissivity}\times ({Ib2}/{Ib1}-1) \times {R_I}$ 
\State $\texttt{QR(cell.ind)} \leftarrow \texttt{QR(cell.ind)} - \texttt{Absorption}$
\Comment{Dump the residual energy into the initial cell}
\EndProcedure
\EndFor
\EndFor
\end{algorithmic}
\end{algorithm}

\subsection{Verification and validation}
To ensure a correct implementation, the algorithm is first verified and validated for a CPU implementation using a combination of grey and non-grey gases in 1D and 3D. In total, 12 cases are used which are summarized in table~\ref{valid_table}. 
Beside the case names in column 1, the second column shows values of the absorption coefficient $\kappa$ for the grey gas cases (cases 1 to 4), and the names of the non-grey gases H$_2$O and CO$_2$ of cases 5 to 12. The other columns indicate the type of the prescribed temperature distribution (linear, parabolic, etc.), the spatial inhomogeneous dimensions (1D or 3D), the wall emissivities $\varepsilon_w$ and the source used for the verification or the validation. 
Further details are given in the subsequent discussions of the individual cases. 

\begin{table}[b]
\centering
\caption{Description of validation cases}
\scalebox{0.85}{
\begin{tabular}{lccccccc}
\hline
Case& $\kappa$ & Temp. & Dimensions & Domain & $\varepsilon_w$ & Comparison\\
\hline
Case 1 & 1 [m$^{-1}$] & lin1 & 1D & 1 [m] & 1 (all walls) & analytical solution\\
Case 2 & 1 [m$^{-1}$] & parab & 1D & 1 [m] & 1 (all walls) & analytical solution\\
Case 3 & 0.5 [m$^{-1}$] & sin & 3D & 1 [m$^3$] & 1 (all walls) & analytical solution~\cite{sakurai2}\\
Case 4 & 5 [m$^{-1}$] & sin & 3D & 1 [m$^3$] & 1 (all walls)& analytical solution~\cite{sakurai2}\\
Case 5 & H$_2$O & 1000 [K] & 1D & 0.1 [m] & 1 (all walls)& Kim et al. \cite{kimrad} \\ 
Case 6 & H$_2$O & 1000 [K] & 1D & 1 [m] & 1 (all walls)& Kim et al. \cite{kimrad} \\ 
Case 7 & CO$_2$ & lin2 & 1D & 1 [m] & 1 (all walls) & Cherkaoui et al. \cite{cherkaoui} \\ 
Case 8 & CO$_2$ & lin2 & 1D & 1 [m] & 0, 1 & Cherkaoui et al. \cite{cherkaoui} \\ 
Case 9 & CO$_2$ & lin2 & 1D & 1 [m] & 0.1, 0.1 & Cherkaoui et al. \cite{cherkaoui} \\ 
Case 10& H$_2$O & parab & 1D & 1 [m] & 1 (all walls) & Line-by-Line MC  \\ 
Case 11& CO$_2$ & parab & 1D & 1 [m] & 1 (all walls) & Line-by-Line MC  \\ 
Case 12& H$_2$O & 3dimens & 3D & 1 [m$^3$] & 1 (all walls) & Line-by-Line MC  \\ 
\hline
\end{tabular}}
\label{valid_table}
\end{table}

\begin{figure}
\centering
\includegraphics[width=1\textwidth]{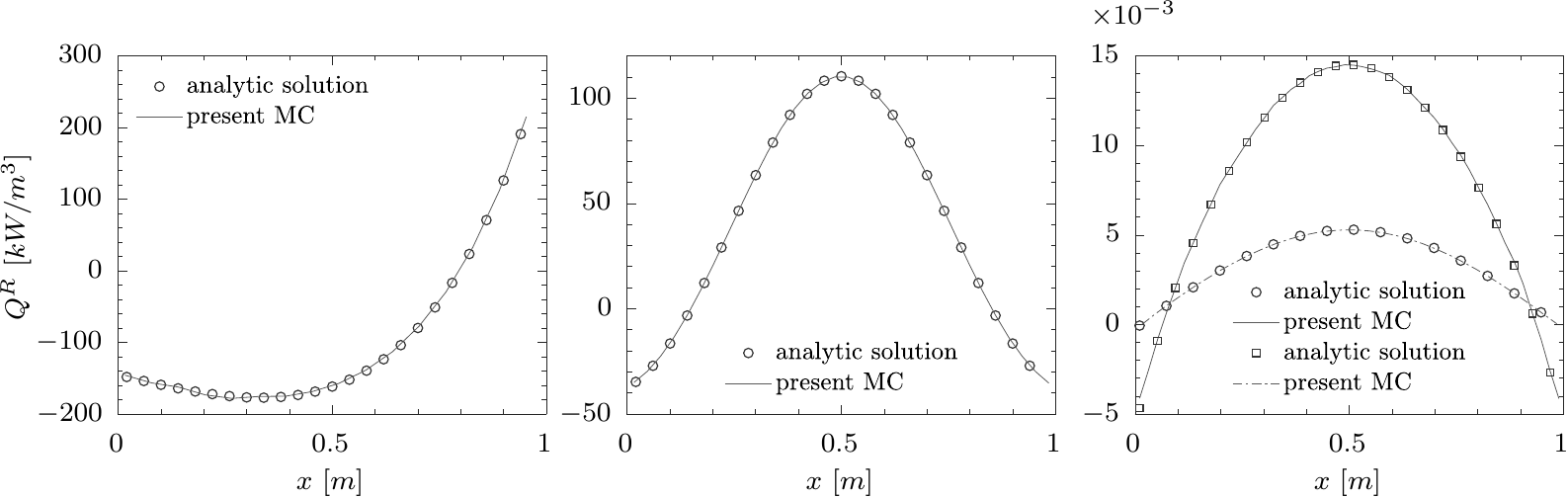}
\caption{Verification of the present MC code (lines) for a grey gas in comparison with analytic solution (symbols). Left: case 1, center: case 2, right: circles and dashed dotted line case 3, squares and solid line case 4 (both at $y = z = 0.5$ [m]).}
\label{grey_valid}
\end{figure}

The grey gas cases 1, 2, 3 and 4 are used to verify the correctness of the ray marching procedure and are compared to existing analytical solutions. Although $\kappa \ne f(\nu)$, the spectral (narrow-band) description shown in algorithm~\ref{euclid} is retained with precomputed probability functions based on a grey gas absorption coefficient. 
Two different geometries are examined, namely a $1$ [m] parallel slab (1D) and a $1$ [m$^3$] cube (3D). 
The walls are considered black with $\varepsilon_w = 1$. 
For the 1D cases, two different temperature profiles (lin1 and parab) are considered, given as 
\begin{alignat}{2}
\label{Tprof}
&\text{lin1:} \  && T_m = 500 + 1000 x \ \text{[K]}, \ \ T_{w1} = 500 \ [K], \ \ T_{w2} = 1500 \ \text{[K]}, \\
\label{Tprof2}
&\text{parab:}  \ \ && T_m = 500 - 2000 x^2 + 2000 x \ \text{[K]}, \ \ T_{w1} = T_{w2} = 500 \ \text{[K]}, 
\end{alignat}
where $T_m$ is the temperature of the medium and $T_{w1}$ and $T_{w2}$ are the temperatures at the left and right wall, respectively. For the 3D cases, the walls are cold ($0$ $[K]$), and the temperature profile is given as  
\begin{equation}
\text{sin:} \ \ T_m = \left(\text{sin}\pi x\cdot\text{sin}\pi y\cdot\text{sin}\pi z\cdot\pi/\sigma  \right)^{0.25} \ [K], 
\end{equation}
in order to compare the results with the quasi-analytic solution derived by Sakurai et al.~\cite{sakurai2}. The absorption coefficient for the 1D slab has a value of $1$ [m$^{-1}$], while for the 3D domain the two cases have different absorption coefficients of $\kappa = 0.5$ and $5$ [m$^{-1}$] (case 3 and 4, respectively). For these four cases, the results are obtained on a $32^3$ grid with $2000$ rays per cell. Note that for the 1D cases, an averaging was performed along the periodic directions. As can be seen in figure~\ref{grey_valid}, the MC implementation is accurately able to reproduce the analytic solutions with adequate precision. 

\begin{figure}[h]
\centering
\includegraphics[width=0.85\textwidth]{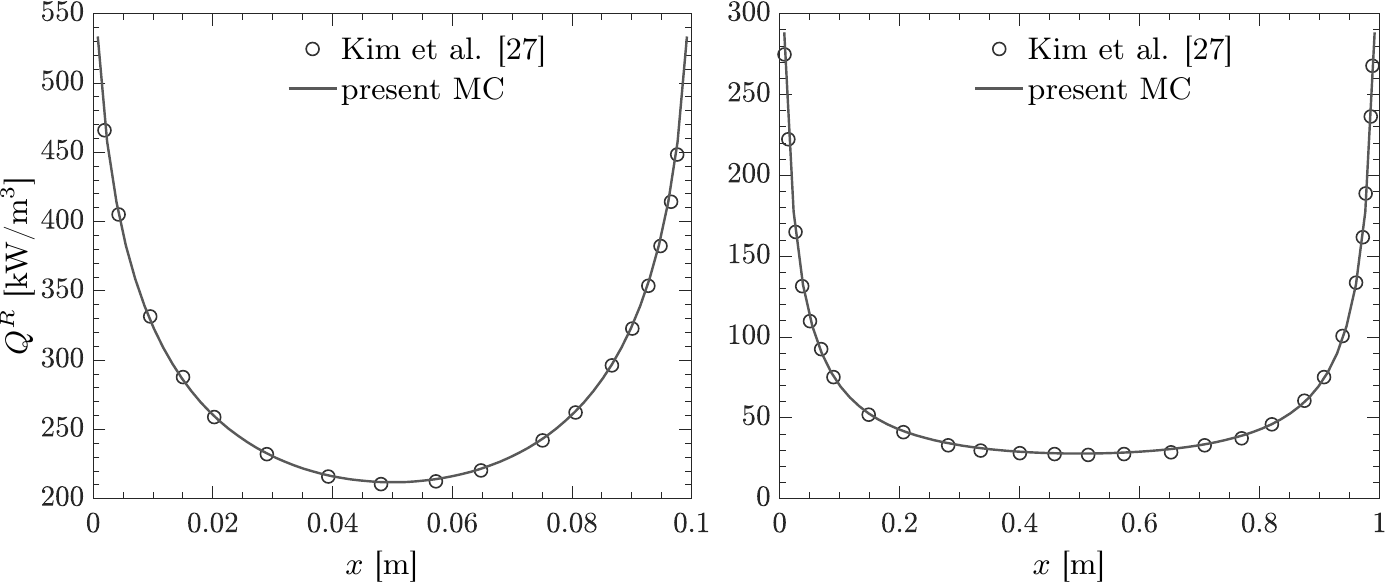}
\caption{Validation of the present MC code (lines) for H$_2$O in the isothermal case in comparison with values from \cite{kimrad} (circles). Left: case 5, right: case 6.}
\label{iso_valid}
\end{figure}

\begin{figure}[h]
\centering
\includegraphics[width=\textwidth]{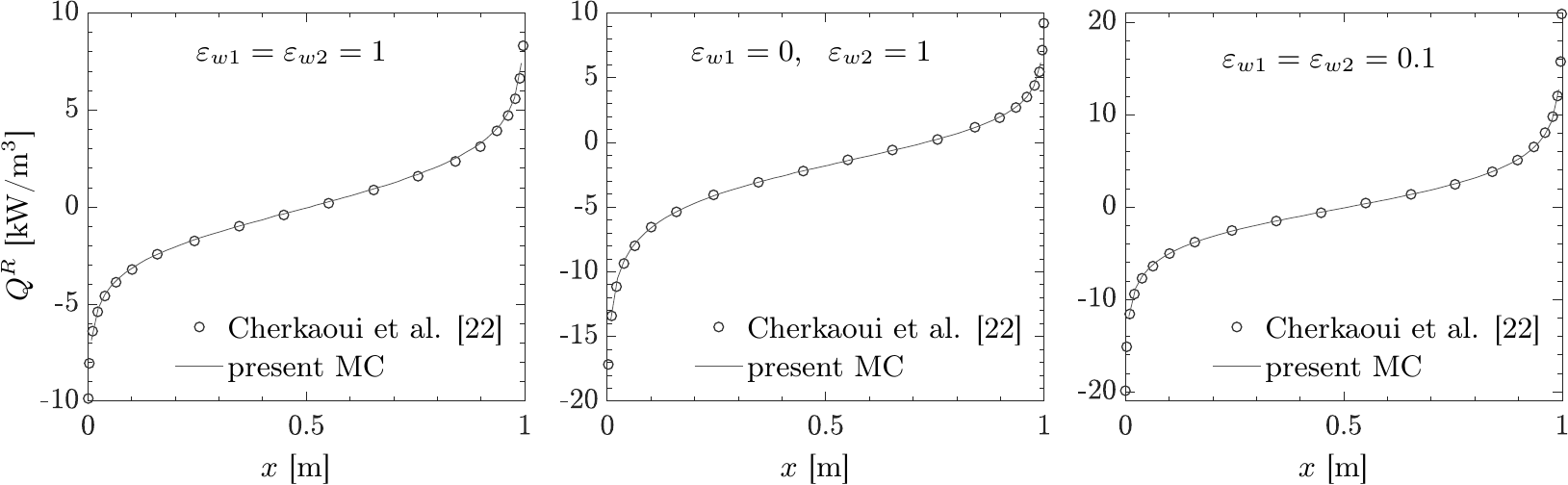}
\caption{Validation of the present MC code for CO$_2$ at 1 [atm] in comparison with results from \cite{cherkaoui}. Linear temperature profile $T=295+10x$ [K]. $T_{w1} =295$ [K] (lin2). $T_{w2} =305$ [K]. Cases 7, 8 and 9 at the left, right and center, respectively.}
\label{CO2_valid}
\end{figure}

To validate the spectral discretization, a combination of isothermal and non-isothermal cases with H$_2$O and CO$_2$ have been used. 119 and 139 narrow bands were selected for H$_2$O and CO$_2$, respectively, with each band containing 16 quadrature points. The radiative power of a 1D slab filled with water vapour at $1$ [atm] and $1000$~[K], bounded by two cold black walls at a distance of $0.1$ (case 5) and $1$ [m] (case 6), has been compared with data presented in Kim et al.~\cite{kimrad} as shown in figure~\ref{iso_valid}. The results for the 1D slab filled with CO$_2$ at 1 [atm] and three different wall emissivities (cases 7, 8 and 9) are shown in figure~\ref{CO2_valid}.  The temperature profiles for the CO$_2$ cases are linear with the left wall at $295$~[K] and the right wall at $305$~[K] (lin2). The radiative power is compared to data presented in Cherkaoui et al.~\cite{cherkaoui}. In all cases (cases 5-9) the comparison clearly demonstrates the high accuracy of the spectral discretization. 


\begin{figure}[h]
\centering
\includegraphics[width=0.9\textwidth]{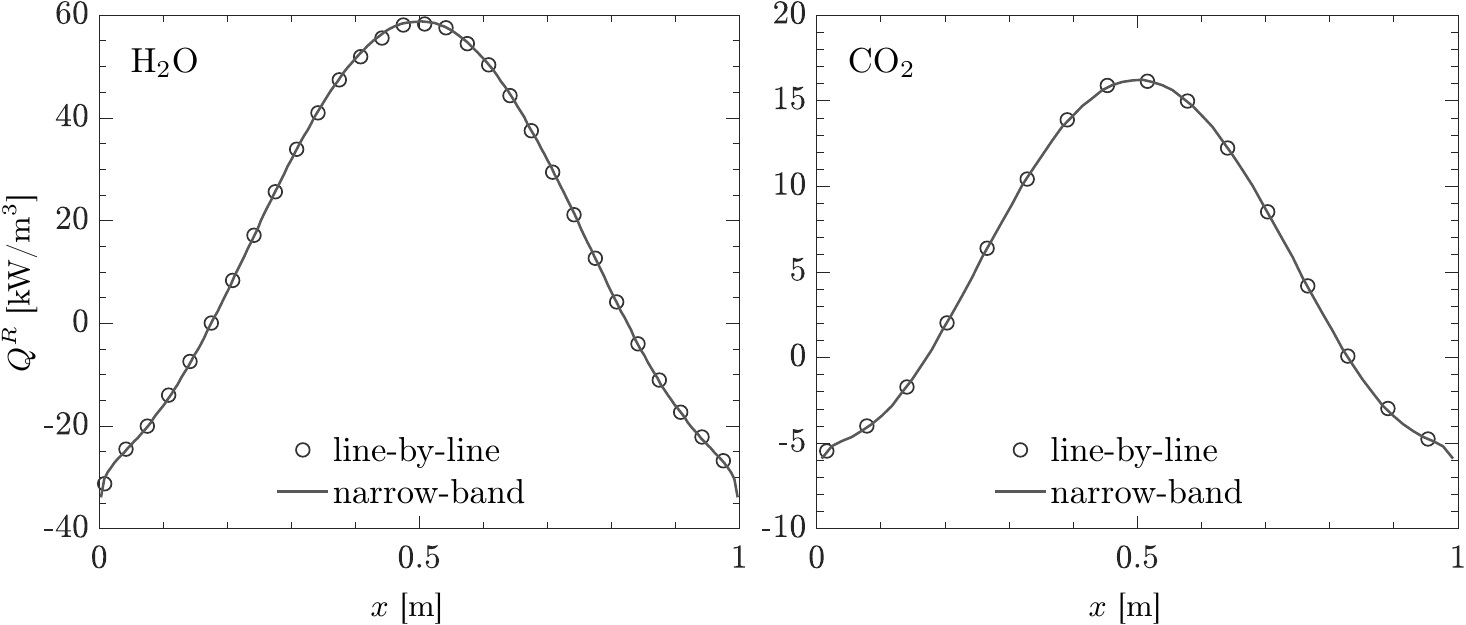}
\caption{Comparison of MC implementation with the line-by-line solution for H20 (left) and CO2 (right) at 1 [atm] with a parabolic temperature profile, equation (\ref{Tprof2}).}
\label{noniso_valid}
\end{figure}

Two additional cases (case 10 and 11) are proposed to validate the spectral discretization and the MC implementation with a line-by-line version of the present MC code. 
The radiative power is calculated for H$_2$O and CO$_2$ at $1$ [atm] with parabolic temperature profiles (equation (\ref{Tprof2})). The results obtained with the narrow-band correlated-$k$ MC, shown in figure~\ref{noniso_valid}, are in close agreement with the line-by-line benchmark to again prove the correct implementation.
 
\begin{figure}[h]
\centering
\includegraphics[width=0.8\textwidth]{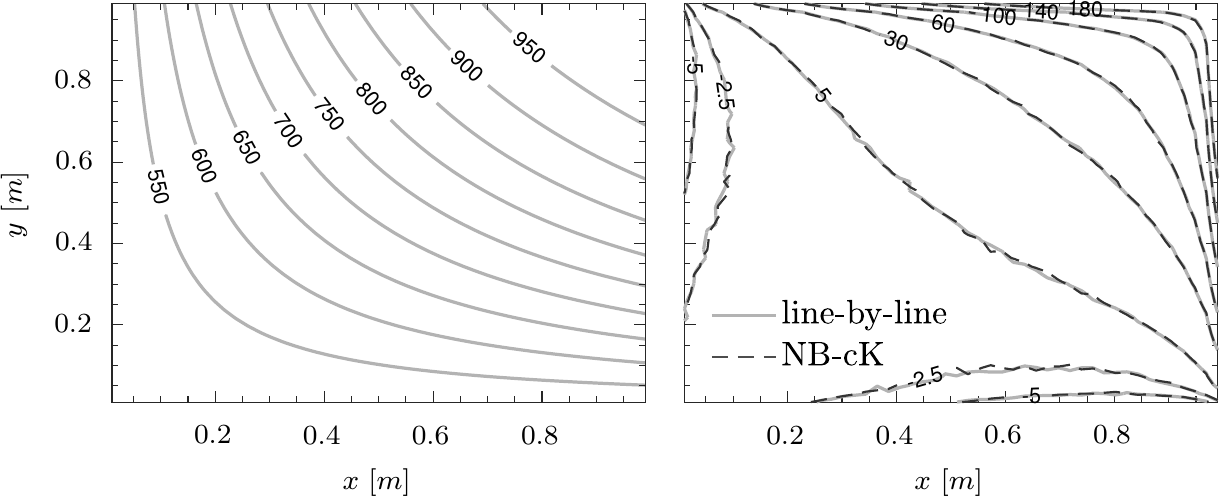}
\caption{3D non grey case. Left: temperature profile at $z = 0.5$ [m] in [K]. Right: radiative power at $z=0.5$ [m] in [kW/m$^3$].}
\label{2D_valid}
\end{figure}

The last validation case (14) consists of 1 [m$^3$] cube with black walls filled with H$_2$O, which is also compared to a line-by-line version of the current MC code. 
 The temperature profile is given by
\begin{equation}
\text{3dimens:} \ \ T_m = 500 - 2000\cdot (x\cdot y\cdot z)^2 + 2000\cdot x\cdot y\cdot z \ [\text{K}], \ \ T_w = 500 \ [\text{K}].
\end{equation}
Figure~\ref{2D_valid} shows the results of the 3D non grey case. The left contour shows the temperature at $z = 0.5$ [m], while the right plot shows the comparison of the results obtained with the narrow-band correlated-$k$ method and the line-by-line benchmark at the same location (shown in [kW/m$^3$]). The solution is again in excellent agreement with the line-by-line benchmark case. 

In the following sections, the 1D H$_2$O parallel slab case with parabolic temperature profile (case 10) will be used to compare the computational performances of the different implementations. Although the case is 1D in nature, it is calculated on a 3D grid with two periodic directions to mimic the computations for a DNS of a fully developed turbulent channel flow. 

\section{GPU implementation} 

Graphical processing units have an architecture that, differently from CPUs, promote compute bound, highly parallelizable algorithms. The smallest parallel GPU units, called threads, run concurrently and are organized in thread blocks. All blocks can read and write into a global memory. The global memory is the \enquote{main} memory of the GPU, comparable to the heap in a C program, and has the slowest I/O access. Threads are grouped into groups of 32, termed \enquote{warps}, which are executed by a single scheduling unit and thus follow a Single Instruction Multiple Thread (SIMT) execution model. Hence, all threads belonging to a particular warp execute the same instruction simultaneously. 
Due to these features, the objective of porting an application from CPU to GPU, is to increase parallelization to favour the SIMT execution. 
A further level of parallelization is obtained by using \enquote{streams}. With this GPU feature, a device function, called \enquote{kernel} can be subdivided into parallel streams that run concurrently and independently, i.e. in a Multiple Instruction Multiple Data (MIMD) fashion, similar to multicore CPU computation (MPI parallelization). In compute bound problems, the use of streams is always recommended, since parallel MIMD execution is preferred to SIMT execution due to the absence of branch divergence (see section \ref{nbsortsection}). 

\begin{figure}
\centering
\includegraphics[width=0.90\textwidth, clip=true]{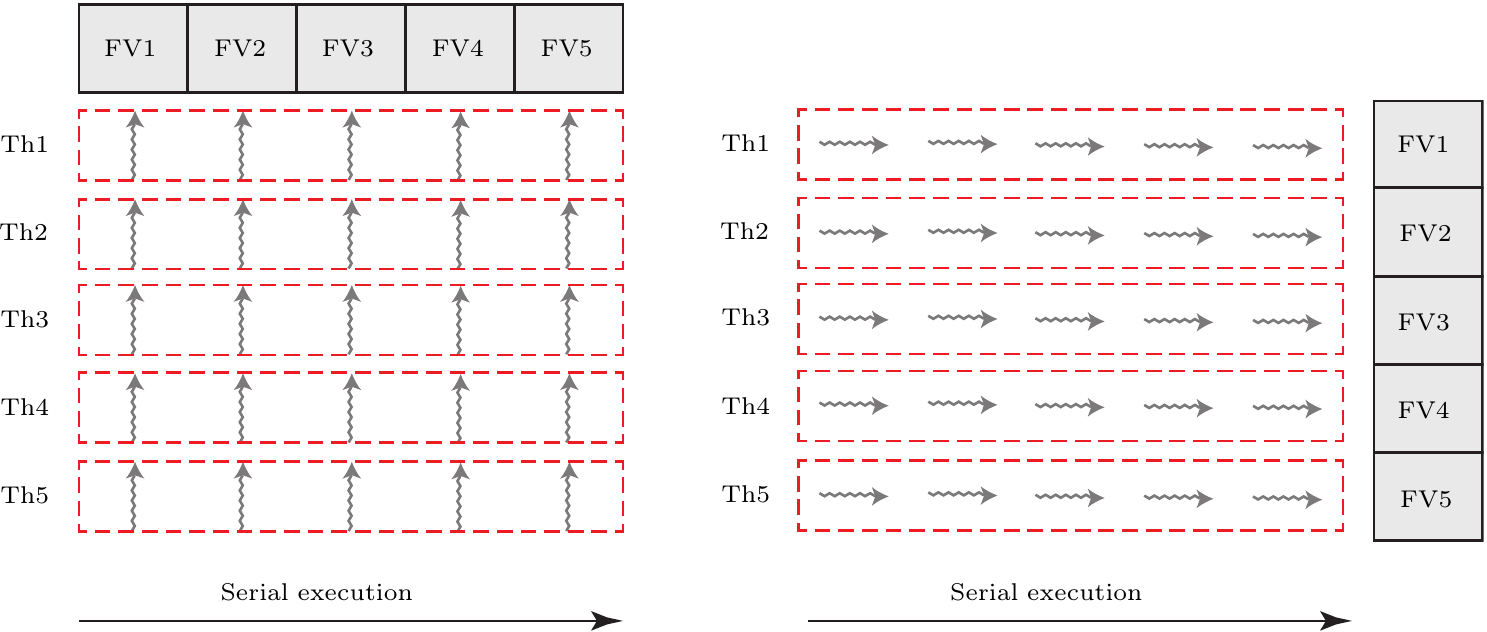}
\caption{Schematic showing the concept of ray parallelization (left) and domain parallelization (right). This simple example is composed of five control volumes ($FV1-FV5$), five rays per cell (grey lines) and five threads working in parallel ($Th1 - Th5$)}
\label{parallelstrat}
\end{figure}

There are two main approaches to parallelize a radiative Monte Carlo algorithm on a GPU. Consider an example with a computational domain of five finite volumes (FV), each one sending five rays to march through the overall domain and five threads (Th) that can execute the marching of the rays. A schematic of this configuration is outlined in figure~\ref{parallelstrat}. The algorithm can then be parallelized by either ray parallelization or domain parallelization, which are outlined in more detail below.

In the first approach, each thread calculates one ray per finite volume. In this case, within each thread the finite volumes are executed in serial, while the rays per cell are parallelized. The solution will, therefore, be obtained by adding the partial results of each thread. The drawback of this approach is the continuous use of expensive atomic reductions (different threads have to read/write in the same memory location). On the other hand, if an ERMC formulation is employed, it is possible to use the second approach, which consists in having a single thread calculate all the rays belonging to an individual finite volume. This is possible due to the fact that, in a reciprocal formulation, the only information required to calculate the radiative source in a point are the rays leaving the latter. In the schematics of figure~\ref{parallelstrat}, the ray decomposition and the domain decomposition approaches are displayed on the left and on the right, respectively. It is important to note that it is also possible to combine the two methods by exploiting the block/thread arrangement. Namely, divide the domain through different blocks and implement a ray parallelization for the finite volumes contained by the block. This configuration would speed up the necessary atomic reductions by enabling the use of shared memory that can be accessed by the whole block. However, shared memory is limited in size and for this reason this approach cannot scale efficiently to larger grids. Given these reasons, we conclude that the domain parallelization approach is more suitable for coupling a GPU Monte Carlo code with DNS. 
 
\begin{algorithm}[t]
\caption{ERMC GPU implementation}\label{euclid2}
\begin{algorithmic}[1]
\scriptsize
\State \texttt{\_\_device\_\_ solution[stream\_max][Ncell/stream\_max]} \Comment{global device variable}
\State \texttt{cudaMemcpyAsync}(Temperature \texttt{T}, absCoeff \texttt{$\kappa$}, Grid, \textit{CopyFromCPUtoGPU}) \Comment{memory copy to device (GPU)}
\For{\texttt{s}$=0$; \texttt{s}$<$\texttt{stream\_max}} \Comment{loop over streams}
        \hspace*{-\fboxsep}\colorbox{red1}{\parbox{\linewidth}{
	\Procedure{kickoff}{thread \texttt{t}, block \texttt{b}, stream \texttt{s}} \Comment{First kernel for stream number \texttt{s}}
		\State $\texttt{\_\_Shared\_\_ state}=\texttt{cuRandInit}$ \Comment{cuRand variable in shared memory}
		\State $\texttt{tid} \leftarrow \texttt{threadIdx.x} + \texttt{blockIdx.x} \times \texttt{blockDim.x}$
		\For{$\texttt{idx}=\texttt{tid}; \texttt{idx}<\texttt{Ncells}; \texttt{idx}=\texttt{idx}+\texttt{blockDim.x}\times\texttt{gridDim.x}$} \Comment{Grid-stride loop over the GPU grid structure}
			\State $\texttt{cell.ind.i} \leftarrow  \texttt{idx} / (\texttt{kmax}\times \texttt{jmax}) + 1 + \texttt{s} \times\texttt{imax}/\texttt{stream\_max}$ 
					\Comment{Mapping thread index to mesh}
			\State $\texttt{cell.ind.j} \leftarrow  \texttt{idx} / \texttt{kmax} + 1 - (\texttt{cell.ind.i}-1-\texttt{s}\times\texttt{imax}/\texttt{stream\_max})\times \texttt{jmax}$
			\State $\texttt{cell.ind.k} \leftarrow  \texttt{idx} - \texttt{kmax}\times (\texttt{cell.ind.j}-1+(\texttt{cell.ind.i}-1-\texttt{s}\times\texttt{imax}/\texttt{stream\_max})\times \texttt{jmax}+1)$
			\State $\texttt{QE} \leftarrow 4\kappa_P(T_{max})\sigma T_{max}^4/$\texttt{numberOfRays}
					\For{ray \textbf{in} Rays}
					\Procedure{Initialize}{}
					\State $R_{\theta},R_{\phi},R_{n},R_{g} \leftarrow$ \texttt{cuRand}(Uniform distribution, \texttt{state}) 
							\Comment{As in the CPU algorithm, but with cuRand instead}
					\State Lines $6-14$ in Algorithm \ref{euclid}
				\EndProcedure
				\Procedure{March}{}
					\State Lines $17-23$ in Algorithm \ref{euclid}
					\State $\texttt{solution[s][idx]} \leftarrow \texttt{solution[s][idx]} -  \texttt{Absorption}$
							\Comment{device global variable that allows asynchronous computations}
					\State Lines $25-30$ in Algorithm \ref{euclid}
				\EndProcedure
			\EndFor
		\EndFor
	\EndProcedure}}
\EndFor
\State Perform other tasks
\For{\texttt{s}$=0$; \texttt{s}$<$\texttt{stream\_max}} \Comment{loop over streams}
        \hspace*{-\fboxsep}\colorbox{red1}{\parbox{\linewidth}{
	\Procedure{return}{thread \texttt{t}, block \texttt{b}, stream \texttt{s}}\Comment{Second kernel for stream number \texttt{s}}
		\State $\texttt{tid} \leftarrow \texttt{threadIdx.x} + \texttt{blockIdx.x} \times \texttt{blockDim.x}$
		\For{$\texttt{idx}=\texttt{tid}; \texttt{idx}<\texttt{Ncells}; \texttt{idx}=\texttt{idx}+\texttt{blockDim.x}\times\texttt{gridDim.x}$}
			\State $\texttt{QR[idx]} \leftarrow \texttt{solution[s][idx]}$
		\EndFor
	\EndProcedure}}
	\State \texttt{cudaMemcpyAsync}(Solution \texttt{QR}, \textit{CopyFromGPUtoCPU}) \Comment{memory copy to host (CPU)}
	\State \texttt{cudaDeviceReset}() \Comment{clear device memory allocations}
\EndFor
\end{algorithmic}
\end{algorithm}

Algorithm \ref{euclid2} displays the GPU implementation of the ERMC based on domain parallelization. The implementation closely resembles the one displayed in algorithm \ref{euclid}, with the difference that the routine now consists of two different GPU functions (or kernels) highlighted in light blue. The first one is in charge of initiating the calculation on the GPU, which immediately returns the control to the CPU, while the second routine retrieves the results. This approach enables a completely asynchronous computation of the GPU and allows to perform other tasks on the CPU (line 26) that would otherwise remain idle. Each kernel is executed \texttt{stream\_max} times and computes (1/\texttt{stream\_max})th of the domain. The stream loops (lines 3 and 27) contain only non-blocking statements that enable a parallel stream execution. The core of the domain parallelization consists in mapping the thread index to a specific finite volume (lines 8-10). The for-loop over the computational cells is then replaced by a GPU-grid-stride loop that runs over the thread index (line 7) and covers all cells in the domain. The random number generation is performed on-the-fly by employing the CUDA library cuRand. The solution is stored in a global device variable \texttt{solution}, which is then retrieved by the second kernel once the computations are complete.    

\begin{table}[b]
\centering
\caption{Comparison between standard CPU and GPU implementation}
\scalebox{0.85}{
\begin{tabular}{lccccccc}
\hline
grid size& $16^3$ & $32^3$ & $48^3$ & $64^3$ & $96^3$ & $128^3$ & $160^3$ \\
\hline
CPU & $    269.4 \ s$ & $   2921.1 \ s$ & $  13182.7 \ s$ & $  39313.3 \ s$ & $ 271844.4 \ s$ & $ (920183.3) \ s$ & $ (2452230.3) \ s$ \\
GPU  & $11.8 \ s$ & $84.8 \ s$ & $394.0 \ s$ & $1169.5 \ s$ & $6143 \ s$ & $19623 \ s$ & $47539 \ s$ \\
Speedup & $     22.8 \times$ & $     34.4 \times$ & $     33.5 \times$ & $     33.6 \times$ & $     44.3 \times$ & $     (46.9) \times$ & $     (51.6) \times$ \\
\hline
rays per cell & $6\cdot 10^2$ & $1.5\cdot 10^3$ & $6\cdot 10^3$ & $1.5\cdot 10^4$ & $3\cdot 10^4$ & $6\cdot 10^4$ & $1.5\cdot 10^5$ \\
\hline
CPU & $    369.7 \ s$ & $    961.7 \ s$ & $   3928.6 \ s$ & $  10432.2 \ s$ & $  19661.1 \ s$ & $  39313.3 \ s$ & $ 132641.3 \ s$ \\
GPU  & $14.1 \ s$ & $31.8\ s$ & $119.2 \ s$ & $294.4 \ s$ & $585.8 \ s$ & $1170 \ s$ & $2916 \ s$ \\
Speedup & $     26.2 \times$ & $     30.2 \times$ & $     33.0 \times$ & $     35.4 \times$ & $     33.6 \times$ & $     33.6 \times$ & $     45.5 \times$ \\
\hline
\end{tabular}}
\label{init_table}
\end{table}

The GPU implementation is tested for case 10 (see table~\ref{valid_table}, plane parallel slab of 1 [atm] H$_2$O with parabolic temperature profile) on a  Tesla K40M. The execution speed is benchmarked against the CPU implementation executed on an Intel Xeon E5-2680 @ 2.40GHz. Table \ref{init_table} shows the computational time required as a function of mesh size and number of rays per cell. In all the test cases, the maximum allowed number of streams (16) is used, while the number of blocks and threads per block are calculated such that the GPU resources are fully utilized. The default values for the parameters that are not varied are a grid size of $64^3$ and $6\cdot 10^4$ rays per cell. The results in table \ref{init_table} show that the speedup obtained with a straightforward GPU implementation is already relatively high. Nonetheless, with the increase of problem size, the speedup does not show a satisfying improvement, reaching values of around $\sim 50 \times$. This apparent limit is caused by the finite resources of the GPU. Being a compute bound algorithm, the scarce resource is the amount of registers per thread that sets the maximum number of threads running concurrently. If the number of registers is increased, the scheduling units serialize the execution of the exceeding warps. As a consequence, no further gain is observed when increasing the mesh size or the number of rays per cells.
Note that the values in parenthesis for the CPU execution time in table~\ref{init_table} are extrapolated from the scaling of the other results and, as such represent an estimation only.

\section{Algorithm acceleration}
A naive GPU implementation, as demonstrated in the section above, is usefull to provide a certain level of speedup, but is certainly not enough to address the computational requirements of a DNS simulation. In particular, the main problems and bottlenecks of such an algorithm are the slow memory access and the large inactivity of the threads due to the SIMT execution model. For this reason, we will address these issues by implementing acceleration techniques that will significantly reduce the execution time and thus enable a full coupling between DNS and the GPU Monte Carlo code. 

\subsection{Texture memory}
\label{textmemory}
Due to the GPU architecture, memory input and output is heavily affected by the access pattern of the threads. In particular, the global memory of a GPU is optimized for coalesced access. A coalesced memory transaction is one in which all of the threads in a half-warp access global memory at the same time. That is to say, consecutive threads should access consecutive memory addresses in the global memory to obtain efficient memory loads/stores. To avoid penalties associated with uncoalesced transactions, it is possible to store variables in registers (the memory associated with the single thread) or shared memory, which is fast-access memory common to all threads in a block. Unfortunately, these two memory types are severely limited in size (on a tesla K40M shared memory consists of only $49$ $kB$ per multiprocessor for a total of $\sim 735$~$kB$). Therefore, after all the fast memory resources have been depleted, it is necessary to store the bulk of the variables in the global memory. Since most memory fetches depend on the drawing of random numbers, it is not easily predictable which address consecutive threads might access. As a consequence, coalesced memory transactions are impossible to achieve in a Monte Carlo simulation. An easy way to optimize memory input and output is hence to employ texture memory. Texture memory is a type of read-only memory, which has been developed for graphical applications. Instead of storing variables linearly, as global memory does, texture memory is designed to optimize the spatial locality of memory access. In other words, each point is associated to a coordinate, and the most efficient memory fetch occurs when consecutive threads access adjacent coordinates in the texture memory instead of consecutive addresses. This scenario is much more likely in a domain parallelized Monte Carlo simulation. The input values to access a texure memory location are float coordinates, while the value returned from the memory is a linear (or trilinear in case of a 3D texture) interpolation of the adjacent values. This feature is extremely useful as it provides fast linear interpolation, which is repeatedly required in a spectral MC code (lines 14, 15, 23 and 25 in algorithm \ref{euclid}). 

\begin{table}[t]
\centering
\caption{Execution time with classical versus textured memory approach.}
\scalebox{0.9}{
\begin{tabular}{lccccccc}
\hline
grid size& $16^3$ & $32^3$ & $48^3$ & $64^3$ & $96^3$ & $128^3$ & $160^3$ \\
\hline
classic  & $11.8\ s$ & $84.8 \ s$ & $394.0\ s$ & $1169.5 \ s$ & $6143\ s$ & $19623\ s$ & $47539\ s$ \\
texture & $8.5\ s$ & $48.6 \ s$ & $260.0\ s$ & $716.0\ s$ & $4381\ s$ & $12343\ s$ & $34047\ s$ \\
Speedup & $1.38 \times$ & $1.74\times$ & $1.52\times$ & $1.64\times$ & $1.40\times$ & $1.59\times$ & $1.40 \times$\\
\hline
rays per cell & $6\cdot 10^2$ & $1.5\cdot 10^3$ & $6\cdot 10^3$ & $1.5\cdot 10^4$ & $3\cdot 10^4$ & $6\cdot 10^4$ & $1.5\cdot 10^5$ \\
\hline
classic  & $ 14.1\ s$ & $31.8 \ s$ & $ 119.2\ s$ & $ 294.4\ s$ & $585.7 \ s$ & $1170  \ s$ & $2916 \ s$ \\
texture  & $ 9.6 \ s$ & $20.6 \ s$ & $ 74.3 \ s$ & $ 180.7\ s$ & $358.6 \ s$ & $714.5 \ s$ & $1784 \ s$ \\
Speedup & $1.47 \times$ & $1.54\times$ & $1.60\times$ & $1.63\times$ & $1.63\times$ & $1.64\times$ & $1.64\times$\\
\hline
\end{tabular}}
\label{texture}
\end{table}
Variables that were residing in the global memory (temperature, blackbody intensity and absorption coefficient), are therefore relocated to the texture memory. The results of the texture memory implementation are shown in table~\ref{texture} in comparison to a standard GPU implementation. The use of texture memory results in a computational gain for all the different settings. Nonetheless, the speedup tends to decrease with mesh size. This behaviour could be caused by the reduced spatial locality of memory access for contiguous threads on a finer grid (i.e. the ray travels further, distancing itself from the aligned source cells). On the other hand, the speedup increases if more memory transactions are performed (i.e., increasing the numbers of rays per cell) 

\begin{figure}[h]
\centering
\includegraphics[width=0.9\textwidth, clip=true]{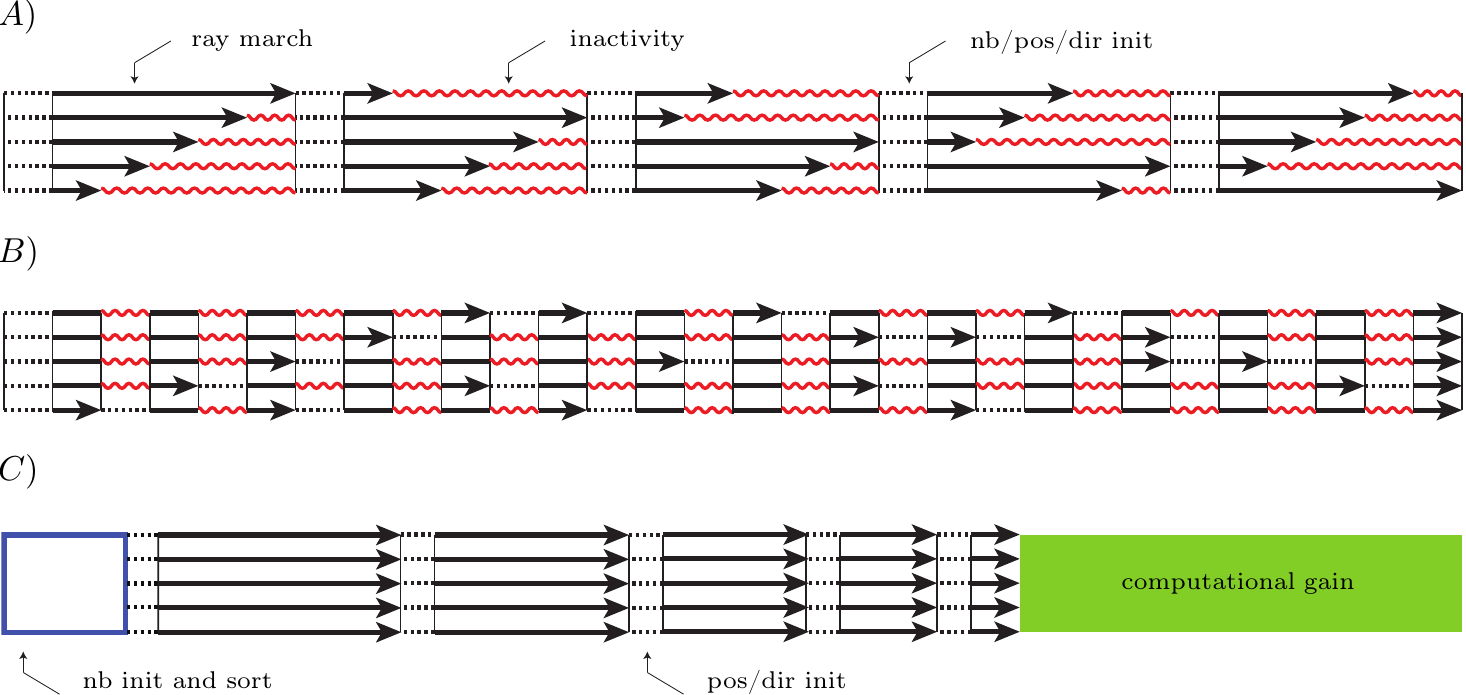}
\caption{Example of a marching procedure for different GPU MC schemes. $(A)$: standard MC implementation; $(B)$: reinitialization MC; $(C)$: sorting MC. The different rows represent the sequential execution of different threads in a warp. We show here only 5 threads and 5 rays to simplify the scheme, but in reality there are 32 threads in a warp and tens of thousand rays per thread. Note that the length of the arrows and the dashed lines (representing marching and initialization) are always preserved among the three schemes. On the other hand, the position and direction initialization time (dashed lines) is shorter in the last scheme ($C$), since the wavelength has already been chosen in the preprocessing step (blue box). }
\label{nbscheme}
\end{figure}

\subsection{Narrow band sorting}
\label{nbsortsection}
The SIMT execution model can lead to a severe performance loss, known as \enquote{branch divergence}. A warp executes one common instruction at a time, so full efficiency is realized when all 32 threads of a warp follow the same execution path. If threads of a warp diverge due to a data-dependent conditional branch, the warp executes all the paths entirely, disabling threads that are not on that path. For the purpose of correctness, the SIMT execution model can be essentially ignored, however, in terms of code efficiency, thread divergence is a serious issue and has to be addresses if the goal is to optimize the algorithm. 

A simple and straightforward approach to reduce inactivity, would be to re-initialize the ray whenever a marching is terminated within the warp. On the other hand, re-initializing the ray on a particular thread forces to temporarily disable the threads that have not yet completed the marching, serializing the initialization procedure. As a consequence, the execution time of multiple initializations might become longer than the benefit obtained by the lower inactivity during the marching procedure. 

\begin{figure}[t]
\centering
\includegraphics[width=0.9\textwidth]{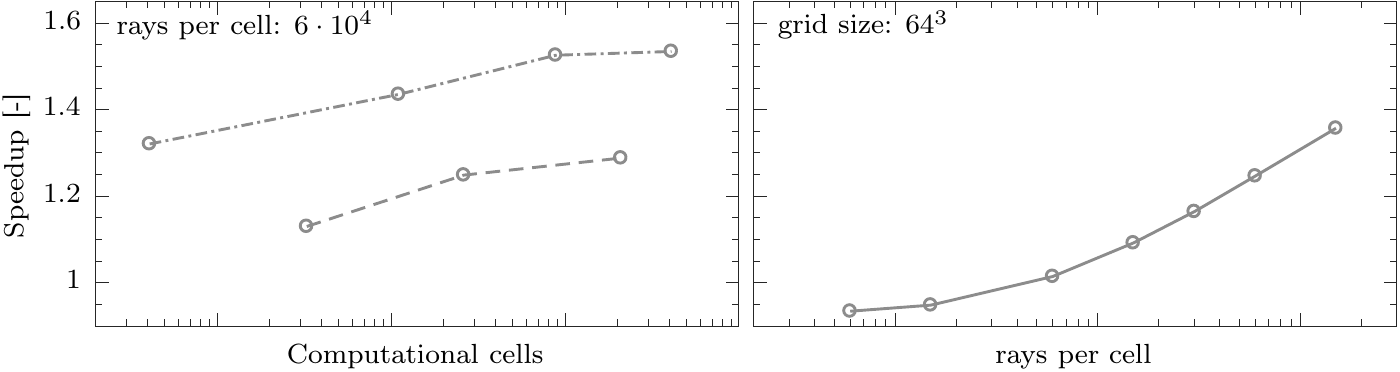}
\caption{Speedup obtained with the narrow-band sorting technique. In the figure on the left, the dashed line connects the points characterized by a mesh which is $2^n\cdot$streams ($16^3$, $32^3$ and $64^3$), while the dashed dotted line all the other points}
\label{speednb}
\end{figure}

Taking into account the properties of the ray, leads to a more effective solution. For example, when two different threads in the same warp are marching rays with different wavelength, they handle different absorption coefficients. 
The ray with a higher $\kappa_{\nu}$ will complete the marching quicker than the one with lower $\kappa_{\nu}$, due to the shorter path length. Since a Monte Carlo routine requires random draws of the wavelength based on a probability distribution function, it is a common scenario that threads are handling absorption coefficients of different order of magnitude. Due to the SIMT execution model, the time required for the warp to complete the current ray tracing is dictated by the thread with the lowest absorption coefficient. It is therefore beneficial to have threads handling absorption coefficient of similar value at all times, such that the tracing might complete simultaneously. To achieve this, it is necessary to precompute all wavelengths for each ray in each finite volume and sort them based on their magnitude of $\kappa_{\nu}$. Consequently, threads will always march rays from the lowest to the highest $\kappa_{\nu}$. While these values might be slightly different for different threads, the order of magnitude of $\kappa_{\nu}$ will be similar, thus significantly reducing the branch divergence of the warp. 

The different configurations are outlined in figure \ref{nbscheme}. The first scheme is a standard MC that does not account for any branch divergence reduction technique. Scheme $B$ shows a re-inizialization scheme in which, wherever a thread in the warp completes the marching, the ray is immediately re-initialized. It is clear that this scheme is successful only if the cost of initializing a ray is smaller than the tracing of the shortest ray. This is not the case in a medium with a high absorption, where rays can be terminated within 5 steps. Scheme $C$ shows the advantage of reordering the rays based on their absorption coefficient which aligns the ray marching executions.

\begin{table}
\centering
\caption{Speedup using the narrow band sorting. The values of the speedup are referred to the textured execution times of table \ref{texture}}
\scalebox{0.9}{
\begin{tabular}{lccccccc}
\hline
grid size& $16^3$ & $32^3$ & $48^3$ & $64^3$ & $96^3$ & $128^3$ & $160^3$ \\
\hline
Time    & $6.4 \ s$     & $43.2 \ s$    & $180.8 \ s$   & $573.0 \ s$   & $2866 \ s$    & $9597 \ s$     & $ 22189 \ s    $ \\
Speedup & $1.32 \times$ & $1.13 \times$ & $1.44 \times$ & $1.25 \times$ & $1.52 \times$ & $1.29 \times $ & $ 1.53 \times $ \\
\hline
rays per cell & $6\cdot 10^2$ & $1.5\cdot 10^3$ & $6\cdot 10^3$ & $1.5\cdot 10^4$ & $3\cdot 10^4$ & $6\cdot 10^4$ & $1.5\cdot 10^5$ \\
\hline
Time    & $10.3 \ s$ & $ 21.7 \ s$ & $73.2 \ s$ & $165.5 \ s$ & $308.0 \ s $ & $573.2 \ s $ & $1313 \ s $ \\
Speedup & $0.93 \times $ & $0.95 \times $ & $1.01 \times  $ & $1.1 \times  $ & $1.16 \times  $ & $1.25 \times  $ & $1.36 \times  $ \\
\hline
\end{tabular}}
\label{sorting}
\end{table}
The results of the tests for a narrow band sorting algorithm are shown in table \ref{sorting} and figure \ref{speednb}. The speedup obtained with sorting the narrow band is larger when the grid is not $2^n \cdot $streams (32, 64, 128). This is caused by an inefficient mapping of the grid onto the device resources, which in this case are powers of 2. Indeed by sorting the narrow-band, it is possible to correct the penalties associated with an inadequate mapping. 
It is possible to notice that the speedup increases with increasing the number of mesh points, until it reaches a plateau for large mesh sizes. On the other hand, if the number of rays per cell are too small, the advantage of a lower warp inactivity is overshadowed by the cost of the sorting procedure. Contrarily, increasing the number of rays per cell leads to an linear growth of the speedup, since the warp inactivity is efficiently replaced by the ray marching computation.

It is interesting to notice the difference between the speedup of the narrow band sorting scheme with respect to mesh size and the speedup using a texture memory approach only. While the first one increases, the latter decreases with grid size. This difference shows the interplay between memory transactions and computations as the mesh size increases, highlighting the larger relative importance of compute statements with increasing mesh size.

\subsection{Multigrid}
\definecolor{grey}{rgb}{0.35, 0.35, 0.35} 
\begin{figure}[hb]
\centering
\includegraphics[width=0.85\textwidth, clip=true]{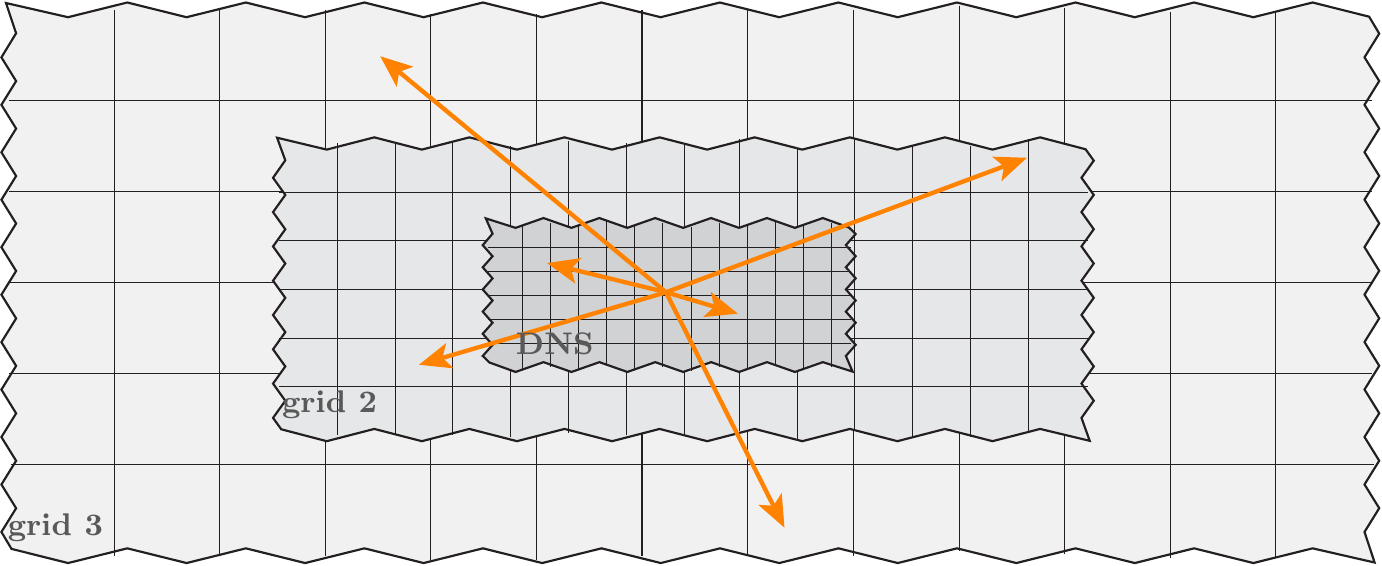}
\caption{Schematic showing the concept of the mesh coarsening scheme. The orange lines symbolize the marched rays. Several grids are overlayed one on top of each other. The ray falls onto the coarsened mesh when it reaches the maximum number of steps in the current grid. The concept is shown here in two dimensions for simplicity.}
\label{adapmesh}
\end{figure}
The radiative intensity is absorbed exponentially as function of the absorption coefficient and the travelled distance. Therefore, the intensity absorbed by traversing a cube of size $\Delta x^3$ will be roughly proportional to
\begin{linenomath*}\begin{equation}
I_{abs} \sim (1-\exp{(-\kappa_{\nu} C \Delta x)} ) \ .
\end{equation}\end{linenomath*}
Consequently, the intensity of the ray leaving the cell is
\begin{linenomath*}\begin{equation}
I_{out} = I_{in} - I_{abs} \sim \exp{(-\kappa_{\nu} C \Delta x)} \ ,
\end{equation}\end{linenomath*}
which signifies that, for a low $\kappa_{\nu}$, the intensity gradient of the propagating ray will be mild and the required cell size $\Delta x$ can be relatively large. Vice versa, if $\kappa_{\nu}$ is large, a lower $\Delta x$ is necessary to capture the steep intensity gradient. 
If an adequate $\Delta x$ is chosen as a pre-processing step (as it could be done in a grey gas medium) the mesh will be over-resolved for the rays with low absorption, resulting in an inefficient ray tracing. Nevertheless, since a high $\kappa_{\nu}$ ray will be terminated fairly quickly, it requires a high resolution only on a small zone around the source point. On the contrary, a ray with low absorption will propagate far into the domain. By combining these two features of rays with different absorption coefficient, it is possible to construct a mesh strategy that optimizes the ray tracing, while retaining a high accuracy. The objective is to have a grid that is fine close to the starting cell and gradually coarser as the ray travels further away from the initial point. To obtain this effect, it is possible to overlay several meshes characterized by different cell sizes. The temperature values will be interpolated on the coarser meshes from the DNS solution which represents the finest mesh level (radiative heat transfer does not introduce new spatial wavenumbers, so the smallest radiative length scales are as small as the Batchelor scales). For all finite volumes, the ray tracing commences on the DNS mesh and the ray is allowed to step onto the current mesh a fixed number of times. If the ray is not exhausted, it falls into a coarser mesh and so forth, until the last mesh is reached. The last (and coarsest) mesh will trace the ray until depletion. The only added overhead is the cost of the interpolation onto coarser meshes, which is completely irrelevant compared to the gain in computational speed obtained. A similar method, involving patches of interest, was previously implemented by Humphrey et al. in two different occasions. Namely, in a parallel CPU Monte Carlo implementation \cite{humphrey2} and in a grey gas GPU implementation \cite{humphrey}. They used this technique to reduce computational and communication time. On the other hand, we highlight the efficiency that such a method has in a non-grey GPU implementation, where it is possible to tailor the method for a pure reduction of thread inactivity caused by the computational mismatch of low and high absorption coefficient's ray execution. Therefore, by targeting the resolution of the higher impact, high absorption rays, the solution retains its accuracy and the parallel efficiency is greatly enhanced.

\begin{figure}
\centering
\includegraphics[width=0.75\textwidth]{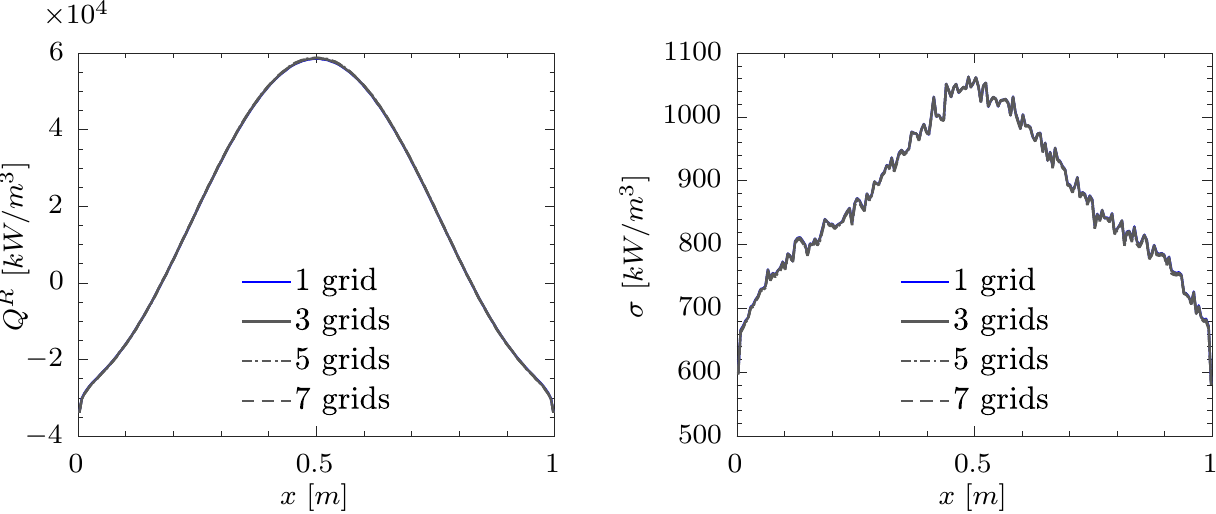}
\caption{Comparison of the results for the parabolic H$_2$O case for different numbers of overlayed coarsened mesh used. 5 steps are allowed in each mesh. $a)$: Radiative heat source, $b)$: standard deviation}
\label{gridadapt}
\end{figure}

Figure \ref{adapmesh} shows a 2D representation of the mesh coarsening concept, while figure \ref{gridadapt} shows the solutions of a test case employing the multigrid technique. In particular, the results shown in figure~\ref{gridadapt} have been obtained with a maximum of $7$ overlayed grids corresponding to $192^3 \rightarrow 96^3 \rightarrow 48^3 \rightarrow 24^3 \rightarrow 12^3 \rightarrow 6^3 \rightarrow 3^3$. The rays were allowed to travel a maximum of $5$ steps in each grid, while proceeding until termination on the last one. It is important to notice that the number of steps has to be accurately decided based on the optical thickness of the grid cells to target the resolution of the relevant $\kappa_{\nu}$. As shown in figure~\ref{gridadapt}, the results of are unaffected by a well tuned grid coarsening technique, both in terms of results and standard deviation. \\
The speedup obtained, defined as $t_1/t_n$, where $t_1$ is the time required for completing the calculation with one grid while $t_n$ with using $n$ grids, is shown in table \ref{speedgrid}. By employing the multigrid technique, it is possible to reduce the computational cost by a factor which is roughly equal to the number of grids used. 
\begin{table}[h]
\centering
\caption{Speedup using multiple overlayed grids. 5 steps per grid}
\scalebox{0.9}{
\begin{tabular}{lccccccc}
\hline
grid number& $1$ & $2$ & $3$ & $4$ & $5$ & $6$ & $7$ \\
Speedup & $1\times$ & $1.4 \times$ & $2.6 \times$ & $4.2 \times$ & $5.8 \times$ & $6.6 \times $ & $ 7.1 \times $ \\
\hline
\end{tabular}}
\label{speedgrid}
\end{table}

\section{Overall performance increase}
\begin{figure}[t]
\centering
\includegraphics[width=\textwidth]{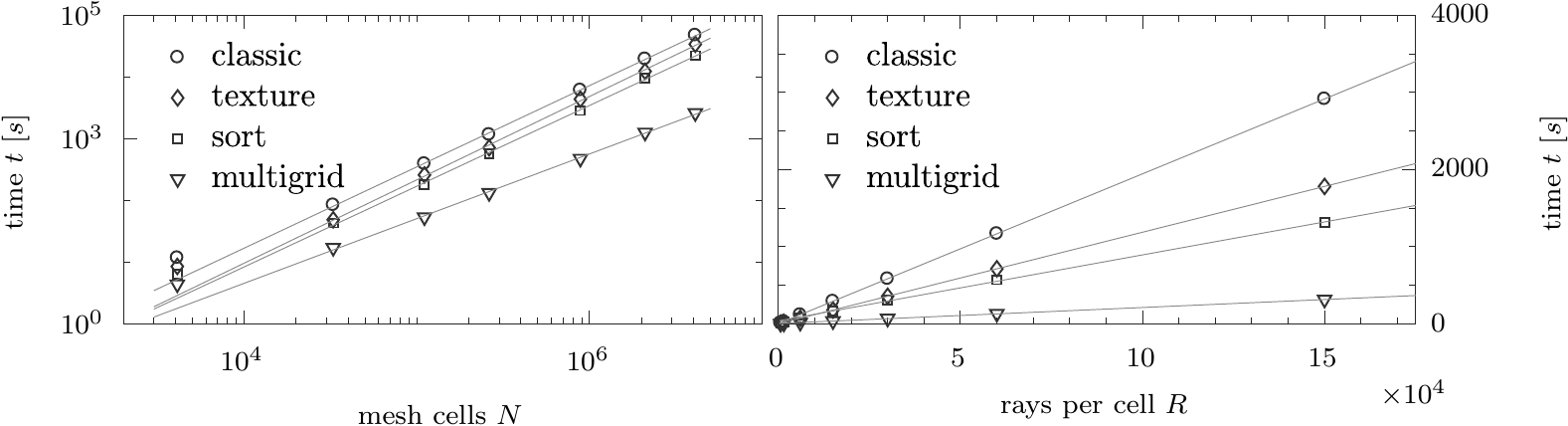}
\caption{Scaling of the code with grid cells and rays per cell.}
\label{final1GPU}
\end{figure}

An overview of the scaling performance using different acceleration techniques is given in figure~\ref{final1GPU} for varying problem sizes. Note that the implementations are additive (i.e., sorting employs texture memory allocations and multigrid performs also a narrow-band sorting). A coarsening ratio of 2 has been employed for successive grids in the multigrid implementation. The smallest allowed mesh had a size of $3^3$, resulting in 3 grids for $16^3$, $4$ for $32^3$, $5$ for $48^3$ and $64^3$ and $6$ grids for $96^3$, $128^3$ and $160^3$. Again, only 5 steps were allowed in each level. The scaling of all implementations is well described by power functions of mesh cells $N$ and linear functions of the number of rays $R$. The grey lines depicted in figure \ref{final1GPU} take the following form 
\begin{itemize}
\item classic $t \propto N^{1.32}$, $t \propto 0.98 R$, 
\item texture $t \propto N^{1.35}$, $t \propto 0.96 R$, 
\item sorting $t \propto N^{1.31}$, $t \propto 0.7 R$, 
\item multigrid $t \propto N^{1.05}$, $t \propto 0.7 R$,
\end{itemize} 
While a texture memory allocation has large benefits for the investigated cases, the computational gain is bound to decrease when the grid size increases (as seen in section \ref{textmemory}) as given by the larger exponent ($1.35 > 1.32$). On the contrary, with a multigrid scheme it is possible to obtain a quasi-linear scaling Monte Carlo code with mesh size ($\text{exponent}\approx1$). Moreover, the narrow band sorting procedure allows a scaling greater than ideal with respect to the rays per cell ($0.7\cdot R$). With more rays being launched, the drawn absorption coefficients fill the whole spectrum space efficiently, replacing the inactivity by aligning more effectively the thread marching.

It is demonstrated that, by employing these optimization techniques, it is possible not only to reduce the computational time, but also to significantly improve the scaling of the code with problem size. The performances of the optimized GPU Monte Carlo code, compared to a serial CPU Monte Carlo implementation executed on an Intel Xeon E5-2680 @ 2.40GHz, is shown in table \ref{final_table}. It has to be reminded that, while texture memory allocation and narrow band sorting only improve computational speed on a GPU, multigrid, although less effective, can be also implemented for a code that runs on a CPU, leading to an increase of code efficiency. The maximum speedup achieved was $570.4\times$ for a grid size of $96^3$. For the largest problems, the CPU computational time was estimated from the scaling. Based on this estimation, we expect a impressive increase of speedup, differently from what is observed in table~\ref{init_table} (potentially we could achieve $938.8\times$ for a $160^3$ grid). 

\begin{table}[h]
\centering
\caption{Comparison between standard CPU implementation and optimized GPU implementation}
\scalebox{0.85}{
\begin{tabular}{lccccccc}
\hline
grid size& $16^3$ & $32^3$ & $48^3$ & $64^3$ & $96^3$ & $128^3$ & $160^3$ \\
\hline
CPU & $    269.4 \ s$ & $   2921.1 \ s$ & $  13182.7 \ s$ & $  39313.3 \ s$ & $ 271844.4 \ s$ & $ (920183.3) \ s$ & $ (2452230.3) \ s$ \\
GPU    & $4.4 \ s$     & $17.1 \ s$    & $53.7 \ s$   & $132.8 \ s$   & $476.6 \ s$    & $1262 \ s$     & $2612 \ s    $ \\
Speedup & $     61.2 \times$ & $    170.8 \times$ & $    245.5 \times$ & $    296.0 \times$ & $    570.4 \times$ & $    (729.1) \times$ & $    (938.8) \times$ \\

\hline
rays per cell & $6\cdot 10^2$ & $1.5\cdot 10^3$ & $6\cdot 10^3$ & $1.5\cdot 10^4$ & $3\cdot 10^4$ & $6\cdot 10^4$ & $1.5\cdot 10^5$ \\
\hline
CPU & $    369.7 \ s$ & $    961.7 \ s$ & $   3928.6 \ s$ & $  10432.2 \ s$ & $  19661.1 \ s$ & $  39313.3 \ s$ & $ 132641.3 \ s$ \\
GPU    & $4.1 \ s$     & $6.3\ s$    & $17.1 \ s$   & $37.5 \ s$   & $69.9 \ s$    & $132.4 \ s$     & $316.0 \ s    $ \\
Speedup & $     90.2 \times$ & $    152.7 \times$ & $    229.7 \times$ & $    278.2 \times$ & $    281.3 \times$ & $    296.9 \times$ & $    419.8 \times$ \\
\hline
\end{tabular}}
\label{final_table}
\end{table}

\section{Multi GPU and DNS coupling}
\begin{figure}
\centering
\includegraphics[width=\textwidth]{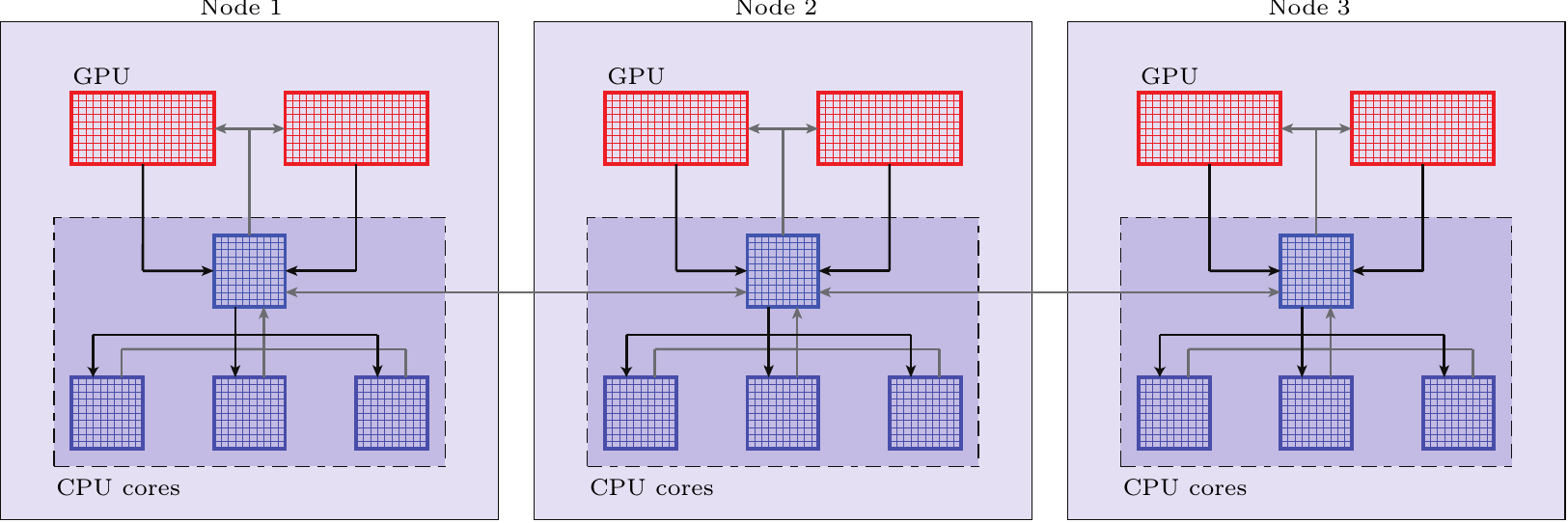}
\caption{Schematic representing the multi GPU implementation. The domain is decomposed on different nodes. In each node one CPU core communicates the the GPUs the entire temperature domain and returns the computed radiative heat source to the CPUs within the node.}
\label{gpuplan}
\end{figure}

The optimized GPU Monte Carlo version can be used to efficiently couple radiative heat transfer with a DNS code in order to study the interactions between radiative heat transfer and turbulent mixing. The coupling is implemented with the use of MPI libraries that handle communications between CPU cores. Each node has a master core which communicates with the available GPUs on the node. Thanks to the reciprocal formulation, the GPUs calculate the radiative source term only on the domain handled by the associated node. On the other hand, to perform ray tracing and to avoid boundary communication, all GPUs require the complete temperature field. A schematic of the multi GPU implementation is shown in figure \ref{gpuplan}. The grey arrows show the communication of the temperature field, while the black arrows show the path of the computed $Q^R$. The memory transfer to and from the GPU is completely asynchronous, such that the CPUs proceed to calculate additional fluid time steps, while the GPUs compute the radiative heat source. As a consequence the CPU computation is completely hidden by the radiative power calculation.

\begin{figure}[h]
\centering
\includegraphics[width=0.9\textwidth]{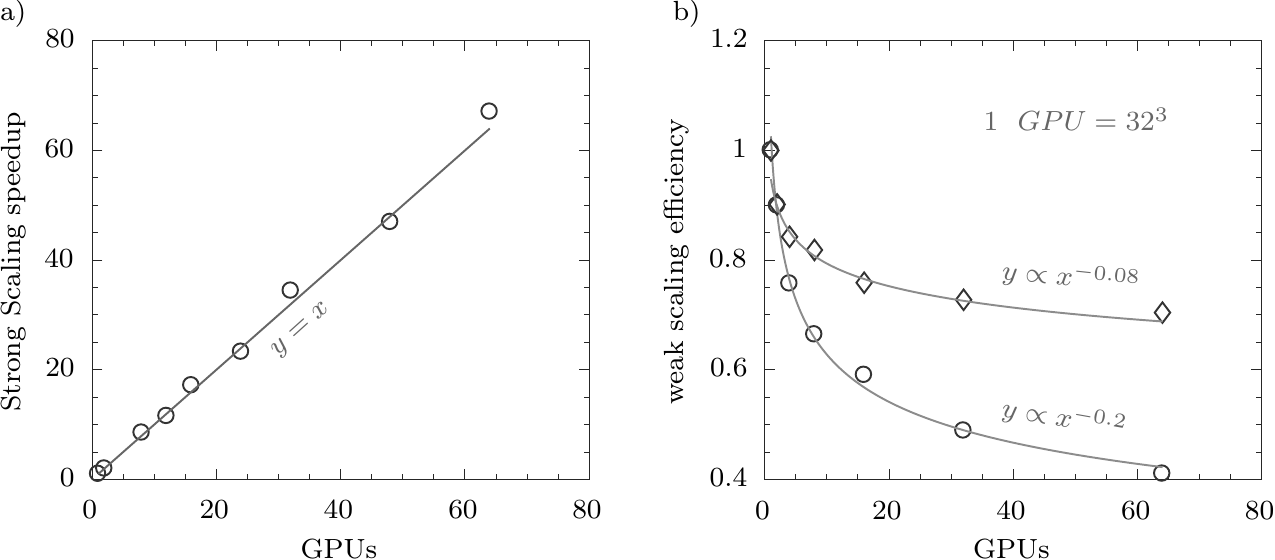}
\caption{Performance of a multi GPU implementation. $a)$: strong scaling speedup ($t_{1}/t_{n}$). $b)$: weak scaling efficiency ($t_{1}/t_{n}$).}
\label{speedmulti}
\end{figure}

The code has been tested on the Cartesius' cluster located in Amsterdam, The Netherlands, on the accelerator island composed of 60 nodes containing 2 Tesla K40M each. The scaling of the code was examined up to 64 GPUs. The results are shown in figure \ref{speedmulti}. The strong scaling of the code is calculated by keeping the grid size constant ($192^3$ in this case) and increasing the number of GPUs. The quantity shown in figure \ref{speedmulti}(a) is the time required for one time step to complete on 1 GPU over the time required for $N$ GPUs. As expected by the computational nature of the code, the scaling is almost ideal. Moreover, figure \ref{speedmulti}(b) shows the weak scaling efficiency, tested with and without the use of the multigrid scheme. In this case the grid size is increased proportionally to the number of GPUs used, with one GPU always computing on a $32^3$ mesh. Since the problem size increases with the number of GPUs used, the code greatly benefits from the multigrid scheme, which improves the weak scaling efficiency from $\propto {GPU}^{-0.2}$ to $\propto {GPU}^{-0.08}$

\begin{figure}
\centering
\includegraphics[width=0.95\textwidth, clip=true]{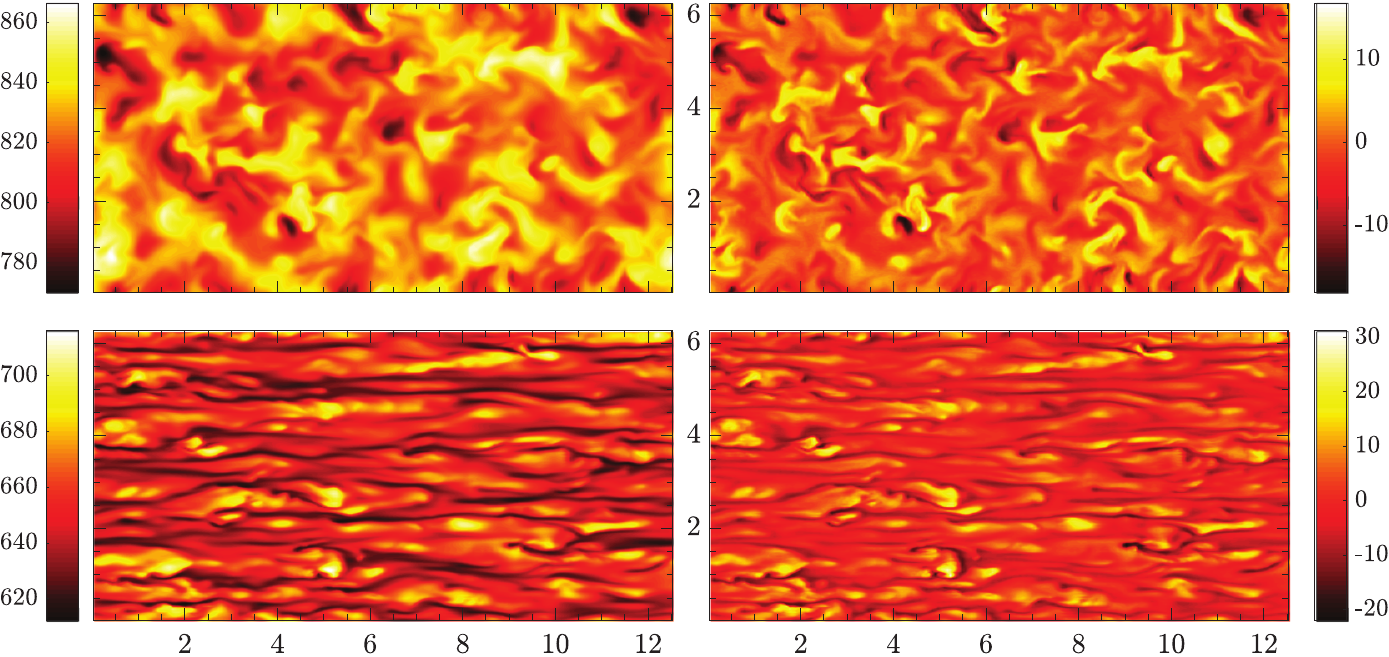}
\caption{Instantaneous snapshots on a wall-parallel plane ($x-z$) at $y = 1.1$ (top) and $y=1.97$ (bottom). Left: temperature $T$ [K]. Right: Radiative power $Q^R$ [kW/m$^3$].}
\label{contjcp}
\end{figure}
\label{sectionres}

To prove the level of accuracy achievable in an acceptable time span, the radiative power is calculated for a turbulent temperature field obtained from a DNS. The DNS represents a fully developed turbulent channel flow with a bulk Reynolds number of $Re=7500$ and isothermal walls at $955$ and $573$ [K] at the bottom and top, respectively. The flow is periodic in the streamwise and spanwise directions. The radiative properties of the medium are those of water vapour at $1$ [atm]. The Planck mean absorption coefficient varies roughly from $5.5$ [m$^{-1}$] near the hot wall to $15$ [m$^{-1}$] near the cold wall and, therefore, can be considered optically thick. In such conditions, the radiative power turbulent spectrum is characterized by short length scales, comparable to the largest wavenumbers of the temperature spectrum. Therefore, the radiative heat source requires to be accurate on the full DNS mesh. The mesh is composed of $192^3$ elements, while the box dimensions are $2$, $2\pi$ and $4\pi$ [m] in the wall normal ($y$), span wise ($z$) and stream wise ($x$) directions, respectively. $6\cdot 10^4$ rays per cells were used to calculate the radiative power. Snapshots of the radiative field are shown in figure \ref{contjcp}. The left contours show the temperature field (in [K]), while the contours on the right are the calculated radiative power in [kW/m$^3$]. The top figures show the fields at a $y$ location of $1.1$ [m] (roughly at the center of the channel), while the bottom figures show a wall normal plane located near the cold wall ($y\approx 1.97$ [m]). As seen from the figures, the radiative field is solved with a high accuracy, matching quite closely the turbulent structures of the temperature field as expected for a highly participating medium. In addition, as predicted in \cite{simone}, in the center of the channel, the turbulent radiative field filters the large turbulent wavenumbers, due to the action of incident radiation acting on the isotropic temperature structures.

\section{Conclusions}
A reciprocal Monte Carlo formulation for radiative heat transfer calculation has been ported to GPU using NVIDIA programming language CUDA. The naive GPU implementation already showed a speedup of almost two order of magnitude compared to a classical CPU implementation. The efforts were focussed on improving the GPU implementation by overcoming the bottlenecks typical of a Monte Carlo code. In particular, the memory access has been enhanced by employing a texture memory for the storage of all read-only variables. This approach allows random memory access and speeds up the computation of the constantly required linear interpolations. Furthermore, The warp inactivity has been significantly reduced using a combination of narrow-band sorting procedures and a multigrid approach. Using this technique the accuracy of the MC solver is retained while the computational expenses are significantly reduced. Therefore, by solving these issues a speedup of up to 3 orders of magnitude when compared to the initial CPU implementation, was achieved. In addition the scaling of the code with problem size (grid cells and rays per cells) was thoroughly studied, demonstrating that the optimized implementation shows a superior scaling when compared to the classical implementation. Indeed the speedup plateau noticed with the standard GPU implementation, is far from being reached even for the largest problems considered.

Moreover, a multi-GPU implementation was performed, showing an efficient strong and weak scaling up to 64 GPUs. While the strong scaling is ideal due to the computational nature of a MC code, the weak scaling benefits largely from the multigrid approach. The coupling with DNS shows the capability of achieving accurate results also for challenging problems as optically thick turbulent flows.

\bibliographystyle{unsrt}
\bibliography{mybibfile}

\end{document}